\documentstyle[aps,eqsecnum,preprint]{revtex}
\begin{document}
\tighten
\draft
%
\preprint{U. of MD PP\#95-123}
\title{Exchange Current Corrections to Neutrino--Nucleus Scattering\\
(I): Nuclear Matter}

\author{Y.~Umino}
\address{Department of Physics, University of Maryland\\
College Park, MD 20742-4111, U.S.A.\footnote{Present address.}\\
and\\
National Institute for Nuclear and High--Energy Physics, 
Section K (NIKHEF--K), \\
Postbus 41882, NL--1009 DB Amsterdam, The Netherlands}
\author{J.M.~Udias}
\address{National Institute for Nuclear and High--Energy Physics, 
Section K (NIKHEF--K), \\
Postbus 41882, NL--1009 DB Amsterdam, The Netherlands}
\date{\today}
\maketitle
\begin{abstract}
Relativistic exchange current corrections to the impulse approximation in 
low and intermediate energy neutrino--nucleus scattering are presented 
assuming non--vanishing strange quark form factors for constituent nucleons.
Two--body exchange current operators which treat {\em all} $SU(3)$ vector and axial 
currents on an equal footing are constructed by generalizing the soft--pion 
dominance method of Chemtob and Rho. 
For charged current reactions, exchange current corrections can reduce the impulse 
approximation results by 5 to 10 \% depending on the nuclear density.
A finite strange quark form factor may change the total cross section for neutral 
current scattering by 20\% while exchange current corrections are found 
to be sensitive to the nuclear density. Implications on the current LSND 
experiment to extract the strange quark axial form factor of the nucleon 
are discussed.
\end{abstract}
%
%
\section{Introduction}
\label{intro}

The existence of two--body meson exchange currents (MEC) in electromagnetic 
and weak axial interactions in nuclei has by now been firmly established \cite{rho79,ris89,fro89}. 
In the electromagnetic sector the need for two--body currents has been 
realized even before the discovery of the pion by Siegert in 1937 through 
his consideration of vector current conservation \cite{sie37}. However, 
it took almost 35 years before the first solid quantitative evidence for MEC was discovered 
when the 10\% discrepancy between theory and experiment in the threshold radiative 
{\em np} capture rates was remedied in terms of one--pion exchange 
current correction \cite{ris72}. The explanation soon after of the cross section for the inverse 
reaction, the electrodisintegration of the deuteron, in terms of MEC 
\cite{hoc73} left little doubt of the important role that two--body currents play in 
electromagnetic interactions of the deuteron. For larger nuclei, it is 
well known that effects of MEC are best found in low and
intermediate energy magnetic isovector processes. These two--body 
currents play important roles in realistic descriptions of diverse nuclear phenomena such as the 
renormalization of orbital $g$--factors, $\delta g_l$ \cite{fuj71}, 
magnetic form factors of light $p$-shell nuclei \cite{dub76}, transverse $(e, e')$ 
response functions in the dip region \cite{ord81,dek91,dek93} and cross--sections 
for electromagnetically induced two--nucleon emission reactions \cite{ryc94}.  

The role of MEC in weak axial transitions in nuclei 
has been investigated by Kubodera, Delorme and Rho \cite{kub78}
who predicted a large renormalization of axial charge due to two--body 
currents. An ideal place to test their prediction is in the 
first--order forbidden $\beta$--decays whose transition amplitude
involves a cancellation between time and space components of the one--body 
nucleon current, thus making it sensitive to two--body MEC 
effects. Indeed, Guichon, Giffon and Samour \cite{gui78} found that the 
impulse approximation prediction of the ratio of $\mu$--capture to
$\beta$--decay rates for the $0^+ \leftrightarrow 0^-$ 
transition between $^{16}O$ and $^{16}N$ was a factor of 2 
larger than the measured value. This large discrepancy disappeared when they 
included two--body MEC corrections to the impulse approximation.
In addition, recent shell model analysis of first forbidden $\beta$-decay 
rates covering a wide range of nuclei \cite{war91,war92,war94} indicates
a substantial exchange current contribution to the renormalization of weak axial 
charge in nuclear medium, thus confirming the prediction of Kubodera, Delorme and Rho.
These solid empirical evidences of two--body currents in electromagnetic and
weak axial processes in low and intermediate energy nuclear phenomena strongly
motivate to investigate whether charged and neutral currents, 
where both vector and axial currents are involved, are also subject to 
renormalizations in nuclei due to MEC.

Another reason to study MEC effects in low and intermediate energy neutrino--nucleus scattering 
is that it has been receiving increasing attention as a means to measure the strangeness 
matrix elements of the nucleon \cite{ahr87,kap88,suz90,hen91,gar92,kol92,gar93,hor93}. 
The measurement of polarized structure function 
$g_1$ and the extraction of the sum rule suggested the possibility of a 
rather large strange quark axial matrix element for the proton leading 
to the so--called ``spin crisis" \cite{ash88}. Although there are numerous works attempting to understand 
the role of hidden flavor in nucleons \cite{ell94}, the situation regarding the  
strangeness degrees of freedom in the nucleon is far from clear and 
it is hoped that neutral current neutrino--nucleus interactions might be able to 
shed a new light into this problem \cite{mus94}.

For example, Garvey {\it et al.} pointed out that the ratio of 
proton--to--neutron yield in quasi--elastic neutral current neutrino--nucleus 
scattering, hereafter denoted by $R(p/n) \equiv \sigma(\nu,\nu' p)/\sigma(\nu,\nu' n)$, 
is an observable that is sensitive to the strange axial form factor of the nucleon \cite{gar92,gar93}. 
This ratio is currently being measured in the LSND experiment at Los Alamos \cite{gar92}. 
In their work Garvey {\it et al.} calculated $R(p/n)$ within a non-relativistic RPA 
framework using the impulse approximation and later included the effects of final 
state interactions experienced by the ejected nucleon with a continuum RPA
formalism \cite{gar93}. This correction was found to have about 
30\% effect on individual neutrino--nucleus cross sections but cancelled out in the ratio of 
proton--to--neutron yields. A similar calculation by Horowitz {\it et al.} \cite{hor93} using a 
Relativistic Fermi Gas (RFG) model in the impulse approximation did not include the final state 
interactions but the resulting $R(p/n)$ was found to be similar to that of Garvey {\it et al.}
However, in order to extract strangeness matrix elements of the nucleon from 
neutral current neutrino--nucleus scattering it is necessary to investigate the 
reliability of calculating the cross sections for this process by 
assuming finite strange quark form factors for the constituent nucleons.
Previously, neutral and charged current neutrino--nucleus scattering have been investigated
beyond the impulse approximation in \cite{min84} but without strange quark form factors 
and MEC corrections.
In addition, this and other proposals \cite{suz90,hen91} 
to extract strange quark form factors of the nucleon from 
neutrino--nucleus scattering involves kinematics ranging from low--energy inelastic scattering to 
the quasi-elastic region. Experience from electron scattering suggests that MEC corrections to 
neutrino--nucleus cross sections in this kinematic range might be important.

The same LSND collaboration has recently announced their measurement
of the cross section for the inclusive charged current reaction 
$^{12}C(\nu_{\mu^-},\mu^-)X$ near threshold 
\cite{alb94} which is significantly lower than existing model predictions. 
This discrepancy is in some cases more than 100\% \cite{kol94} and it is conjectured that nuclear 
effects which are important in low--energy neutrino--nucleus interactions 
have been left out in these model calculations. One such effect is the MEC corrections 
to the impulse approximation and it is therefore interesting to explore whether two--body corrections 
to the one--body nucleon current will help to remedy this recently observed discrepancy. 

Previously, MEC corrections have been investigated only in neutrino--deuteron reactions 
using non--relativistic kinematics and {\em without} assuming any strange quark form factors 
\cite{mul81,sin86,bac88,tat90}.
In a recent letter \cite{umi95} a method to construct relativistic MEC operators applicable 
to neutrino--nucleus scattering was presented taking into account the 
possible finite strange quark form factors of nucleons. This method 
treats all the $SU(3)$ currents on the same footing and thus is able to estimate MEC corrections to 
electromagnetic as well as neutral and charged current processes simultaneously. 
It is also model independent in the sense that no underlying nucleon--nucleon 
interaction needs to be specified in order to construct the MEC operators. 
In this paper details of the method is presented together with examples 
of the use of the resulting MEC operators in
electron and neutrino scatterings assuming nuclear matter and using the 
kinematics of the on--going LSND experiment. 
The focus of this paper is to investigate only the effects of exchange 
currents, and therefore a simple RFG model is used 
to model the target nucleus. In order to make a more realistic comparison with 
experiment additional nuclear effects must be incorporated in the 
description of the neutrino--nucleus scattering process. These effects will be considered 
in a forthcoming paper \cite{udi95}.
  
In the following section the problem is defined and empirical and theoretical motivations are 
presented for the use of the method originally developed by Chemtob and Rho
to construct MEC operators \cite{che71}. Their formalism is then 
generalized to take into account finite strange quark 
form factors and advantages of using the generalized method and
assumptions made in constructing the MEC operators are discussed. 
Section III illustrates the usefulness of the generalized operators 
by evaluating two--body MEC corrections to quasi--elastic 
electron scattering as well as neutral and charged current neutrino 
scattering reactions. Finally, the results are summarized in the concluding section 
accompanied by Appendicies which present all the necessary formalism needed to 
evaluate the nuclear response functions in neutrino--nucleus 
scattering together with some new technical details. 

\section{Exchange Current Operators for All Currents}

\subsection{Currents and Form Factors}
This section presents the construction of MEC operators applicable to 
both electron and neutrino--nucleus scattering by generalizing the method 
developed by Chemtob and Rho \cite{che71}. As will be shown below, this 
generalized method is able to estimate exchange current corrections to {\em any} linear 
combination of $SU(3)$ vector and axial currents given by
\begin{equation}
J_\mu (k)= \sum_{a=0}^{8} \Bigl( \alpha^a V_\mu^a(k) + \beta^a A_\mu^a(k) \Bigr)
\label{eq:GENCUR}
\end{equation}
where $V^a_{\mu}$ and $A^a_{\mu}$ are the $SU(3)$ vector 
and axial vector currents of the nucleon, respectively, and $k \equiv 
k_\mu$ is the four momentum of the incoming probe. 
The $SU(3)$ singlet ($a=0$) and octet ($a=1 \rightarrow 8$) currents are 
defined using the usual Pauli, $F_1^a(Q^2)$, Dirac, 
$F_2^a(Q^2)$, axial, $G_A^a(Q^2)$, and induced 
axial $H_A^a(Q^2)$ {\em on--shell} nucleon form factors where $Q^2 \equiv 
-k^2$
\begin{eqnarray}
V^0_{\mu}(k) & \equiv & \sqrt{\frac{2}{3}} \Bigl[ F_1^0(Q^2)
\gamma_{\mu} + F_2^0(Q^2)\Sigma_{\mu}\Bigr] {\bf I} 
\label{eq:VSING} \\
V^i_{\mu}(k) & \equiv & \Bigl[ F_1^i(Q^2)\gamma_{\mu} + 
F_2^i(Q^2) \Sigma_{\mu} \Bigr] \lambda^i 
\label{eq:VOCT} \\
A^0_{\mu}(k) & \equiv & \sqrt{\frac{2}{3}} \Bigl[ G_A^0(Q^2)
\gamma_{\mu} \gamma_5 + H_A^0(Q^2) k_{\mu}\gamma_5 \Bigr] {\bf I} 
\label{eq:ASING} \\
A^i_{\mu}(k) & \equiv & \Bigl[ G_A^i(Q^2)\gamma_{\mu}\gamma_5 + 
H_A^i(Q^2)k_{\mu}\gamma_5 \Bigr] \lambda^i
\label{eq:AOCT}
\end{eqnarray}
Here $i = 1 \rightarrow 8$, $\lambda^i$ are the usual Gell-Mann matrices
normalized to ${\rm Tr}\:(\lambda^a \lambda^b) = 2\delta_{ab}$, ${\bf I}$ 
is the identity matrix and the magnetic operator $\Sigma_\mu$ is defined as
\begin{equation}
\Sigma_\mu \equiv \frac{i}{2M}\sigma_{\mu\nu}k^\nu
\end{equation}
with $M$ being the free nucleon mass.
Thus, the problem addressed in this paper is to estimate two--body MEC corrections 
to the general one--body $SU(3)$ nucleon current shown in 
Eq.~(\ref{eq:GENCUR}) when the nucleon is immersed 
in nuclear medium. Once this is accomplished, MEC operators 
corresponding to electromagnetic, weak axial, neutral and charged 
currents may be trivially constructed by taking the appropriate linear 
combinations. 

For example, the one--body nucleon electromagnetic current is recovered 
when $\alpha^3 = 1$ and $\alpha^8 = 1/\sqrt{3}$ and by letting the remaining 
coefficients vanish, multiplied by the appropriate isoscalar and 
isovector electric charges, {\em i.e.}
\begin{equation}
J_\mu^{EM}  \propto \left(V_\mu^3+ \frac{1}{\sqrt{3}} V_\mu^8 \right)
\label{eq:EMCUR}
\end{equation}
Similarly, one--body neutral, $J^{Z^0}_{\mu}$ and charged, $J^{W^\pm}_\mu$, 
currents are given by the following linear combinations of vector and axial currents \cite{kap88}
\begin{eqnarray}
J^{Z^0}_{\mu}  & \propto & V^3_{\mu} - A^3_{\mu} - 2\, \sin^2\theta_W \left( V^3_{\mu} + 
\frac{1}{\sqrt{3}} V^8_{\mu} \right) - \frac{1}{2} \left( V^0_{\mu} - 
\frac{2}{\sqrt{3}}V^8_{\mu} \right)
+ \frac{1}{2} \left( A^0_{\mu} - \frac{2}{\sqrt{3}}A^8_{\mu} \right)
\label{eq:ZZERO} \\
J^{W^\pm}_\mu & \propto & \Biggl[ \Bigl( V^1_\mu \pm i V^2_\mu \Bigr)
- \Bigl( A^1_\mu \pm i A^2_\mu \Bigr) \Biggr] \cos\theta_C 
+ \Biggl[ \Bigl( V^4_\mu \pm i V^5_\mu \Bigr)
- \Bigl( A^4_\mu \pm i A^5_\mu \Bigr) \Biggr] \sin\theta_C
\label{eq:JPM}
\end{eqnarray}
In these definitions of neutral and charged currents, $\theta_W$ and 
$\theta_C$ are the Weinberg and Cabbibo angles, respectively, and small 
QED, QCD and heavy quark corrections to $J^{Z^0}_{\mu}$ \cite{kap88}
as well as contributions from the charmed quarks to $J^{W^\pm}_\mu$ are 
ignored. Note that the third term in Eq.~(\ref{eq:ZZERO}) is proportional to the electromagnetic 
current, Eq.~(\ref{eq:EMCUR}), while the last two terms are referred to as the strange vector, 
$V^s_{\mu}$, and axial, $A^s_{\mu}$, currents of the nucleon.
\begin{eqnarray}
V^s_{\mu} & \equiv & V^0_{\mu} - \frac{2}{\sqrt{3}}V^8_{\mu} 
\label{eq:VSTRANGE}\\
A^s_{\mu} & \equiv & A^0_{\mu} - \frac{2}{\sqrt{3}}A^8_{\mu} 
\label{eq:ASTRANGE}
\end{eqnarray}

For neutral current processes where only massless leptons are involved, 
the induced axial form factor does not contribute to the total cross section and
therefore is not determined. At $Q^2=0$, the strange quark Pauli form factor 
$F_1^s \equiv F_1^0 - 2/\sqrt{3}F_1^8$ vanishes by definition and only $F_2^0$ and 
$G_A^0$ are unknown among the form factors in Eqs.~(\ref{eq:VSING})--(\ref{eq:AOCT}). 
According to the Standard Model, these two form factors determine the strange quark 
magnetic, $F_2^s \equiv F_2^s(0)$, and axial, $G_A^s \equiv G_A^s(0)$, form factors of the nucleon.
In this work all form factor parametrizations are taken from 
\cite{bei90} which are the same ones used in \cite{gar92}.
Specifically, the $Q^2$ dependence of $F_2^s(Q^2)$ and $G_A^s(Q^2)$ are 
\begin{eqnarray}
F_2^s(Q^2)  & \equiv & F_2^0(Q^2) - \frac{2}{\sqrt{3}}F_2^8(Q^2) \\
G_A^s(Q^2) & \equiv & G_A^0(Q^2) - \frac{2}{\sqrt{3}}G_A^8(Q^2)
\end{eqnarray}
where
\begin{eqnarray}
F_2^{0,8}(Q^2) & \equiv &  
\frac{F_2^{0,8}(0)}{(1+\frac{Q^2}{4M_N^2})(1+\frac{Q^2}{M_V^2})^2} \\
G_A^{0,8}(Q^2) & \equiv & 
\frac{G_A^{0,8}(0)}{(1+\frac{Q^2}{M_A^2})^2}
\end{eqnarray}
In these definitions the vector and axial masses are set to $M_V$ = 840 MeV and $M_A$ = 1030 MeV, 
respectively, and the octet form factors at $Q^2=0$ are $F_2^8(0) 
\equiv \sqrt{3}/2(\kappa_p + \kappa_n)$ and $G_A^8(0) \equiv \sqrt{3}/6(3F -D)$. 

\subsection{Soft--Pion Exchange Dominance and the Chiral Filter Hypothesis}
Although empirical evidences abound suggesting that both one--body electromagnetic 
and weak axial currents are renormalized in nuclear medium by MEC, there still 
lacks a rigorous theoretical understanding of the roles that different types of MEC might 
play in nuclei. For example, in electromagnetic processes any conserved transverse 
MEC consistent with a given $N-N$ interaction is acceptable since it is not constrained by the 
Ward--Takahashi identity. A typical ``brute force" approach of estimating 
MEC effects in electro--nuclear phenomena is to choose a model dependent $N-N$ 
interaction in the one--boson exchange approximation and from this 
interaction construct the MEC operators with a longitudinal component 
satisfying the Ward--Takahashi identity and a corresponding conserved transverse component. 
In addition, in applications to many--body systems it is necessary to 
construct nuclear wave functions {\em from the same} $N-N$ interaction 
including the Fock term to maintain self--consistency. This latter requirement is often neglected 
without any justifications especially in relativistic calculations. The situation in the weak axial 
case is more ambiguous since there are no conservation laws to constrain the form of 
MEC operators except in the chiral limit.
Thus, it is desirable to identify the important contributions 
from a multitude of two--body currents involving exchanges of different types of mesons 
by exploiting some underlying physical principles. 
Fortunately, there has been some promising experimental and theoretical progress over the 
past 20 years towards accomplishing this goal.

In 1978, Kubodera, Delorme and Rho \cite{kub78}, using the method developed by 
Chemtob and Rho to be described below, have argued that, {\em in the absence 
of kinematical suppressions}, MEC processes 
in low and intermediate energy nuclear phenomena are dominated by one--pion exchange 
whose production amplitude is evaluated in the soft--pion limit. In another words, they 
stressed that dominant contribution to the two--body correction arises from
MEC which are consistent with nuclear force implied by chiral symmetry, other 
short--ranged meson exchanges being ``filtered out" by the nuclear medium. 
Thus, the crucial element in their argument is the 
important role played by chiral symmetry in nuclei, a key symmetry manifest in QCD.  
Using this so--called {\em chiral filter hypothesis}, they arrived at the prediction of axial 
charge renormalization which subsequently was confirmed through the 
analysis of first forbidden $\beta$--decay rates as mentioned in the introductory section. 
In general, Kubodera, Delorme and Rho found that MEC derived assuming 
soft--pion dominance approximation, hereafter referred to as soft--pion 
MEC, strongly renormalizes the time component of axial 
currents and the space component of electromagnetic currents, respectively. 

The idea of soft--pion dominance has been applied in the past to various 
low and intermediate energy phenomena and proved to be a viable 
technique in estimating exchange current corrections.
For example, in their analysis of first forbidden $\beta$--decay rates 
Warburton {\em et al.} compared the use of the soft--pion exchange dominance 
approximation to other MEC operators and found that
both methods can reproduce the observed enhancement of the axial charge\cite{war92,war94}. 
The successful application of 
soft--pion MEC operators here is not surprising since $\beta$--decays 
typically involve small momentum transfers where the soft--pion dominance
approximation is expected to apply. The real surprise came when the
success of this approximation manifested itself in the electromagnetic sector. 
Initially, Riska and Brown \cite{ris72} 
used the soft--pion MEC operators as prescribed by Chemtob and Rho to explain
the difference between the impulse approximation prediction and the 
measured threshold radiative $np$ capture rates.  
The same method was employed by Hockert {\em et al.} \cite{hoc73} to 
calculate the cross section for the 
electrodisintegration of the deuteron which involved small energy 
($E_{np} \approx 3$MeV) but {\em large} momentum transfers. They 
were able to reproduce the measured cross section up to momentum transfer 
of $k_\mu^2 = 10$ ${\rm fm^{-2}}$ using only the soft--pion MEC operators. 
Addition of the $\Delta$ resonance contribution had little effect on 
their original correction to the impulse approximation. The 
same cross section was later measured at Saclay \cite{saclay} extending the 
momentum transfer up to $k_\mu^2 = 18$ ${\rm fm^{-2}}$. The surprise came when 
the original prediction by Hockert {\em et al.} managed to reproduce the 
Saclay data up to $k_\mu^2 = 15$ ${\rm fm^{-2}}$ \cite{rho81}. 
In this case different corrections to the soft--pion MEC  
cancelled each other leaving the original two--body contribution to be the 
dominant correction to the impulse approximation. 
Thus, there are concrete empirical evidences to support the use of the 
soft--pion dominance approximation in low and intermediate energy electromagnetic and 
weak axial interactions in nuclei.

Recently, Rho \cite{rho91} has proposed an explanation for the success of 
the soft--pion MEC dominance based on Weinberg's derivation of nuclear forces from
chiral Lagrangians \cite{wei79,wei90}. Using chiral power counting arguments
Rho has shown that to the leading order, {\em i.e.} at the tree level, 
the short range part of two--body MEC corresponding to a 
nuclear force predicted by a given chiral Lagrangian is considerably suppressed. 
Thus, the dominant contribution to the two--body correction to the impulse 
approximation comes from the long ranged part of the exchange current 
represented by the soft--pion exchange.
Subsequently, Park, Towner and Kubodera \cite{par94} have calculated next--to--leading
order corrections to the axial charge MEC operators beyond the soft--pion 
dominance approximation using heavy fermion chiral perturbation theory. They found that  
loop corrections to the soft--pion MEC operators are of the order of 10\%, and
argued that their results are consistent with the claims of Warburton {\em et al.} and support
the chiral filtering conjecture. Thus, not only are there empirical 
evidences suggesting the renormalization of electromagnetic and 
weak axial currents by MEC in nuclei, but there also exists theoretical 
motivations to believe that this renormalization is dominated by 
soft--pion exchange interaction, at least up to momentum transfers of 
about one GeV \cite{rho81,rho91}.

\subsection{Soft--Pion MEC Operators}
The method of Chemtob and Rho to construct soft--pion MEC operators 
\cite{che71} is based on soft--pion theorems and current algebra techniques 
pioneered by Adler \cite{adl68}. Here, this method is generalized to accommodate 
{\em all} the $SU(3)$ currents appearing in Eq.~(\ref{eq:GENCUR}) and the main 
advantage of using this technique to estimate MEC corrections to neutrino--nucleus 
scattering is pointed out. 
Since the derivation of soft--pion MEC operators may be found in the 
original works of Adler \cite{adl68} and Chemtob and Rho \cite{che71}, 
and the generalization to $SU(3)$ being straightforward, 
most of the technical details are relegated to Appendix~B.

In the soft--pion dominance approximation, the operator representing an exchange 
of a pion between two nucleons is written as products of the pion production amplitude by an 
external current off the first nucleon, the pion propagator and the matrix element for pion
absorption by the second nucleon 
\begin{eqnarray}
J_\mu^a(k;P_{I,1};P_{I,2};P_{F,1};P_{F,2})_{EX} & = & \nonumber \\
&    & \!\!\!\!\!\!\!\!\!\!\!\!\!\!\!\!\!\!\!\!\!\!\!\!\!
\frac{1}{(2\pi)^4} \delta^4(P_{I,1} + P_{I,2} + k - P_{F,1} - P_{F,2}) 
\langle N(P_{F,1}) \pi^b(q)| J_{\mu}^a(k) | N(P_{I,1}) \rangle \nonumber \\
&    & \;\;\; \times \frac{i}{q^2 - m_\pi^2 } 
\langle N(P_{F,2})| J_{\pi}^b(q) | N(P_{I,2}) \rangle + (1 \leftrightarrow 2)                                                                                
\label{eq:JEX}
\end{eqnarray}
Here the matrix element $\langle N(P_{F,1}) \pi^b (q)| J_{\mu}^a(k) | N(P_{I,1}) \rangle$ 
is the amplitude for pion production off a nucleon by an external vector 
or axial $SU(3)$ current $J_{\mu}^a(k)$, 
$J_{\mu}^a(k) + N(P_{I,1}) \rightarrow \pi^b(q) + N(P_{F,1})$, 
where $q \equiv q_\mu$ is the four momenta of the produced pion. 
$a$ and $b$ are $SU(3)$ indices with $a = 0 
\rightarrow 8$ to accommodate all the $SU(3)$ currents and $b =$ 1, 2 or 3 for pion production.
Similarly, $\langle N(P_{F,2})| J_{\pi}^b(q) | N(P_{I,2}) \rangle$ is the 
matrix element for pion absorption by a nucleon using the pseudoscalar $\pi NN$ coupling
\begin{equation}
\langle N(P_{F,2})| J_{\pi}^b(q) | N(P_{I,2}) \rangle = 
g_{\pi NN}\; \langle N(P_{F,2})| \gamma_5\lambda^b | N(P_{I,2}) \rangle
\label{eq:PINN}
\end{equation}
{\em Note the absence of $\pi NN$ form factor in Eq.~(\ref{eq:PINN}).} It is quite remarkable 
that the present method can describe the electrodisintegration data 
involving momentum transfers of up to about $k_\mu^2 = 15$ ${\rm fm}^2$ 
without the use of any $\pi NN$ form factors in the $\pi NN$ absorption 
vertex. 

To construct soft--pion MEC operators it is necessary to 
know the pion production amplitude in Eq.~(\ref{eq:JEX}) in the 
soft--pion limit of $q \rightarrow 0$. In this limiting procedure it is necessary 
to first take the spatial part of the four vector to zero in order to select 
the long range, {\em i e.} the S--wave, part of the $N-N$ interaction and 
{\em then} take the chiral limit of $q_0 \rightarrow 0$.
The resulting amplitude, originally derived by Adler \cite{adl68} using the PCAC hypothesis 
\begin{equation}
\partial^{\mu}A^a_{\mu} = m_{\pi}^2 F_{\pi} \pi^a
\end{equation}
and used by Chemtob and Rho in \cite{che71} has the following form when 
generalized to the $SU(3)$ formalism 
\begin{eqnarray}
\lim_{q \rightarrow 0} \langle N(P_F) \pi^b (q)| J_{\mu}^a(k) | N(P_I) \rangle  
& = &
\frac{i}{F_{\pi}} \int d^4 \! x \lim_{q \rightarrow 0} (-i q^{\mu})
\langle N(P_F)| T \left( A_{\mu}^b(x) J_{\mu}^a(0) \right) |N(P_I) \rangle \\
\nonumber
&    &
\;\;\;\;  - \frac{i}{F_{\pi}}  
\langle N(P_F)| \left[ Q^b_5(x), J_{\mu}^a(0) \right]_{x_0=0} | N(P_I) \rangle
\label{eq:LSZ}
\end{eqnarray}
Here $Q^a_5(x) \equiv \int d^3x A_0^a(x)$ is the axial charge. 
As shown in \cite{adl65}, the only contributions to the first term in the 
soft--pion limit come from pole diagrams where the axial current 
$A_{\mu}$ is inserted in the external lines in the amplitude 
$\langle N(P_F) |J_{\mu}^a(k)| N(P_I) \rangle$ and thus behaving as $1/q_\mu$. 
The second term may be simplified by using the $SU(3) \otimes SU(3)$ current algebra
\begin{eqnarray}
\left[ Q^a_5(x), V_{\mu}^b(0) \right]_{x_0=0} = i f_{abc} A_{\mu}^c(0) \\
\left[ Q^a_5(x), A_{\mu}^b(0) \right]_{x_0=0} = i f_{abc} V_{\mu}^c(0)
\end{eqnarray}
and has no contributions from singlet currents unlike in the first term 
where both $SU(3)$ singlet and octet can contribute.
Therefore, the amplitude for soft--pion emission in the reaction 
$N(P_I) \rightarrow N(P_F)$, in the presence of perturbation $J^a_{\mu}(k)$,
may be expressed in terms of two matrix elements 
$\langle N(P_F)| J_{\mu}^a(k) |N(P_I) \rangle$ and 
$\langle N(P_F)| \left[ Q^b_5(x), J_{\mu}^a(0) \right]_{x_0=0} | N(P_I) \rangle$.
Since $J^a_{\mu}$ may be any one of vector or axial $SU(3)$ currents,
Eq.~(\ref{eq:JEX}) may simultaneously be applied to all the components of the 
general one--body $SU(3)$ current in Eq.~(\ref{eq:GENCUR}), and 
specifically to the electromagnetic current of Eq.~(\ref{eq:EMCUR}) as 
well as to neutral and charged currents of Eqs.~(\ref{eq:ZZERO}) and 
(\ref{eq:JPM}), respectively. Thus, it is the use of $SU(3) \otimes SU(3)$ current 
algebra in Eq.~(\ref{eq:LSZ}), which rotates around the vector and axial 
currents, that makes the generalized method of Chemtob and Rho 
particularly suitable to approximate MEC corrections in 
neutrino--nucleus scattering at low and intermediate energies. In 
addition, because the soft--pion limit is taken no $N-N$ interaction needs to be specified 
and the present method of constructing MEC operators only 
requires the currents in the impulse approximation as inputs. 
In this sense the present approach to constructing MEC operators is model independent. 
Note also that the present method is valid {\em to all orders} in $g_{\pi 
NN}$ in one of the $\pi NN$ verticies since the pion production amplitude Eq.~(\ref{eq:LSZ}) 
is evaluated non--perturbatively \cite{adl68}.\footnote{The expressions
for the soft--pion production amplitudes presented in Appendix~B are 
proportional to $g_{\pi NN}$. However, this does not mean that the 
amplitudes are evaluated to first order in $g_{\pi NN}$. The $\pi NN$ coupling constant was 
introduced through the use of the Goldberger--Trieman relation 
\cite{adl68}.} 

It is useful to discuss the approximations made in calculating the soft--pion 
production amplitudes. Since the original application of soft--pion 
theorems has been on pion photo--production processes, these amplitudes have been evaluated by 
assuming that the initial and final nucleons are on their mass shell. 
However, in order to construct MEC operators in nuclei consistently it is 
necessary to take both the initial and the final nucleons off their mass 
shell. This involves density dependent off--shell parametrizations of nucleon currents which 
are not known. In fact, a fully consistent  
many--body description of in--medium nucleon electromagnetic or weak form 
factors has never been presented. Considering this lack of understanding of   
off--shell modification of nucleon form factors and currents in nuclear 
medium, the most reasonable approximation to make 
is the use of on--shell kinematics and parametrizations of nucleon form 
factors as has been done in all previous works involving MEC in nuclei. 
This implies that the nucleons are assumed to obey the free Dirac equation, and 
therefore in the derivation of the pion production amplitude $p\!\!\!/ 
u(\vec{p})$ has been replaced with $M u(\vec{p})$, where $u(\vec{p})$ is the 
nucleon spinor with three momentum $\vec{p}$ and mass $M$.
This on--shell approximation is used from the very beginning, even before considering 
the soft--pion limit, and is consistent with using the {\em free} RFG model of the 
nucleus to calculate the cross sections where the constituent nucleons are assumed to be 
on--shell.

Related to the derivation of soft--pion production amplitudes, Eq.~(\ref{eq:LSZ}),  
is the use of PCAC and pion off--shell extrapolation. When using the 
PCAC relation, it is assumed that the matrix element of the 
divergence of the axial current varies smoothly and {\em slowly} with the 
pion mass $q^2$. In addition it is also assumed that this matrix element 
is approximately proportional to the pion field and that any dependence 
on higher order non-linear pion field terms are negligible.
This assumption allows one to make a connection between the 
results obtained with $q \approx 0$ with a more realistic value of $q^2$.
Since a zero four momentum pion is not a physical object, this slow $q^2$ 
variation assumption is always needed to compare the soft--pion predictions with experiment.
Finally, the soft--pion production amplitudes are valid only to zeroth order in $q$. This means 
that if there are any processes contributing to the pion production which, 
for kinematical reasons, are of first order in $q$, the applications of 
current algebra and soft-pion techniques used here give no information 
about them. In order to describe these processes, it is necessary 
to make a model dependent analysis of specific reactions which requires 
the introduction of a $N-N$ interaction.

Relativistic $SU(3)$ soft--pion MEC operators are constructed in analogy with the method 
outlined in \cite{che71} and are listed in Appendix~B. 
The conservation of the vector current has been checked analytically 
using the prescription discussed in the original paper by Adler \cite{adl68}
and verified numerically .
To illustrate the usefulness of the present technique, soft--pion MEC corrections are 
applied to quasi--elastic electromagnetic, neutral and charged current 
interactions simultaneously in the following section. 
In this paper a simple RFG model formalism \cite{RFG1,RFG2} is used for all calculations without binding energy corrections ($B=0$).
Finite nucleus effects, final state interactions and other 
density dependent nuclear medium effects are not considered on purpose in order to clearly isolate 
the effects of soft--pion MEC in many body systems. 

\section{Results}

\subsection{Quasi--Elastic Electron Scattering}
As mentioned in Section~IIA, the third term in the expression for the 
one--body neutral current of the nucleon, Eq.~(\ref{eq:ZZERO}), is proportional to the 
electromagnetic current, $J_{\mu}^{EM} \propto V^3_{\mu} + \frac{1}{\sqrt{3}} V^8_{\mu}$. 
Thus, estimates of MEC effects in electron scattering can automatically be 
extracted when calculating two--body corrections to neutral current processes 
using the formalism under consideration. The successful application of the 
soft--pion MEC approximation to the electrodisintegration of the deuteron 
has been described in the previous section. Here, the same method is 
applied to quasi--elastic electron scattering off heavier nuclei where 
many--body effects, not present in the reaction involving the deuteron, are 
expected to play important roles. 

In Figures~1a and 1b, separated 
longitudinal, $R_L(\omega,|\vec{k}\:|)$, and transverse, $R_T(\omega,|\vec{k}\:|)$, 
response functions for a typical quasi--elastic inclusive $^{12}C(e, e')$ reaction 
are shown as a function of energy transfer $\omega$ both in the impulse approximation 
and with soft--MEC corrections. 
A Fermi momentum of $k_F$ = 225 MeV and a fixed three momentum transfer of 
$|\vec{k}\:|$ = 400 MeV are used in the RFG model of the nucleus without 
any binding energy corrections ($B=0$) so that the nucleons in the target 
nucleus are {\em on--shell}.
Since the momentum transfer is almost twice the Fermi momentum, 
the influence of Pauli blocking should be small in this kinematical range and 
indeed, as shown in the figures, the small effect of Pauli blocking is 
manifested in the linear dependence of the response functions for small $\omega$ \cite{RFG1}.
Thus, the RFG model of the nucleus is adequate as a first approximation to test the 
soft--pion dominance approximation. 
Only $1p1h$ final states are considered when evaluating the MEC matrix 
elements since they are the dominant two--body contributions to the response functions in the 
quasi--elastic region.
A more quantitative two--body MEC corrections to electron scattering at 
these energies, and especially in the dip region, would require model dependent 
$\Delta$ propagation and pion production processes as pointed out in \cite{ord81,dek91}.
Also shown in the figures for reference are the experimental data for the same inclusive 
electron scattering reaction measured at Saclay \cite{bar83}. {\em However, no 
attempt has been made to fit the data by varying $k_F$ and $B$,} since 
the vector currents will no longer be conserved with a finite binding 
energy correction and the target nucleons will be off their mass shell.

As shown in Figure~1a, the MEC correction to the impulse approximation in the longitudinal 
response results in a small reduction of the quasi--elastic peak with no 
appreciable change in the peak position.
However, this reduction of the impulse approximation result is too small 
to describe the magnitude of the quasi--elastic peak observed in the Saclay data. 
This feature persists even if $k_F$ and $B$ are varied in an attempt for 
a better fit to the data. The failure of the free Fermi Gas model to describe the observed 
longitudinal response in quasi--elastic inclusive electron scattering for 
a wide range of nuclei is well--known \cite{orl91}, and additional 
nuclear effects must be incorporated in order to improve the present model calculation.
Also shown in the figure is the individual soft--pion MEC contribution 
which is small and negative resulting in the small reduction of the 
quasi--elastic peak.

For the transverse response shown in Figure~1b, the soft--pion MEC 
correction can increase the magnitude of the quasi--elastic peak by about 
20\% relative to the impulse approximation results and shift the peak 
position to a lower value of energy transfer $\omega$ by about 20 MeV.
These effects of the two--body current correction may be understood by 
examining the individual soft--pion MEC contribution which causes both 
constructive and destructive interferences depending on the value of the 
energy transfer. This contribution is positive between $\omega$ = 0 and 
125 MeV and negative thereafter and results in an increase of the 
magnitude of the quasi--elastic peak and a shift of the peak position to a lower 
value of $\omega$. The qualitative agreement with data is better than in 
the longitudinal response although the observed peak position can not be 
reproduced with zero binding energy correction ($B=0$). 
A finite value of the binding energy will help to shift the 
quasi--elastic peak towards the observed position but, as mentioned above, the total 
electromagnetic current will no longer be conserved in this case. 
Thus, two--body soft--pion MEC corrections to inclusive electron scattering response functions 
tend to slightly soften the longitudinal response and leads to about 20\% 
increase in the transverse response in the free RFG model. 

It is interesting to compare the present results with numerous model 
calculations of $R_L$ and $R_T$ in the literature. However, to make any 
quantitative comparisons with other 
works are very difficult since many model dependent assumptions have been made for 
each of the calculations which are hard to disentangle. For example, Kohno and Ohtsuka \cite{koh81}
also investigate the MEC corrections to the inclusive electron 
scattering response functions in the quasi--elastic region considering 
only the $1p1h$ final states. In contrast to the present work, they work in the non--relativistic 
approximation and ignore contributions of order greater than  $g_{\pi 
NN}^2$. The method adopted in this paper is fully relativistic and one of 
the $\pi NN$ verticies is valid to all orders in $g_{\pi NN}$. Also, they
include the $\Delta$ excitation current which has a strong model dependence 
as well as the ``pion--in--flight" MEC, referred to as ``pionic" MEC in 
their paper. As discussed in Appendix~B, the pion--in--flight diagram 
does not contribute to the $(e, e')$ response functions since it is 
proportional to the four momentum transfer of the leptonic probe, $k_\mu$, in the soft--pion limit.
Nevertheless, there is qualitative agreement between the soft--pion MEC 
contribution shown in Figures~1a and 1b and their corresponding ``pion pair" MEC results presented 
in Figure 2b of \cite{koh81}. It is interesting to note that in their calculation 
for the longitudinal response, there is a cancellation between the 
``pionic" and $\Delta$ excitation MEC corrections leaving the ``pair" current 
the dominant correction to the impulse approximation as in Figure~1a of 
the present paper. 
This is not true for their transverse response where Kohno and Ohtsuka  
find that the total effect of the two--body corrections is to decrease 
the impulse approximation result which is the opposite effect shown in Figure~1b. 
The reason for this discrepancy is due to the large negative contribution 
from their $\Delta$ excitation current which is not included in the present work. 
It should be noted that another RFG model calculation of $R_T$ that includes the 
same types of MEC corrections as in Kohno and Ohtsuka \cite{dek91}, but 
considers only the $2p2h$ final states, obtains an increase in the 
transverse response since the $2p2h$ matrix elements are added incoherently to the 
impulse approximation result.
Thus, unlike the case of the deuteron, it is difficult to make a definitive 
statement on the success of the application of soft--pion 
MEC dominance approximation in electron scattering off nuclei by 
comparing with other works, except to say that the 
results presented above are not inconsistent with other model calculations.

\subsection{Quasi--Elastic Charged Current Neutrino Scattering}
The one--body charged current, Eq.~(\ref{eq:JPM}), is the next simplest 
linear combination of $SU(3)$ vector and axial currents of physical interest after the 
electromagnetic current discussed above. Since the nucleon has no {\em 
net} strangeness ($F_1^s(Q^2=0)=0$), only the term proportional to cos$\theta_C$ in 
Eq.~(\ref{eq:JPM}) contributes to the charged current neutrino--nucleus 
scattering cross sections. Although no direct information regarding the 
strange quark form factors of the nucleon may be obtained in charged 
current processes, these reactions are nevertheless important in 
extracting the value of the axial mass parameter $M_A$ which appears in
the parametrization of axial form factors. Recently, the LSND collaboration 
has reported on their measurement of neutrino flux--averaged inclusive 
$^{12}C(\nu_{\mu^-},\mu^-)X$ cross section of (8.3 $\pm$ 0.7 stat. $\pm$ 1.6 
syst.)$\times 10^{-40}\;\;{\rm cm}^2$ in the neutrino energy region of 
$123.7 < E_\nu < 280$ MeV with a flux--weighted average of $<E_\nu>$ 
= 180 MeV \cite{alb94}. This value is substantially smaller than an
earlier measurement, which used a different neutrino energy distribution, 
of (15.9 $\pm$ 2.6 stat. $\pm$ 3.7 syst.)$\times 10^{-39}\;\;{\rm cm}^2$ 
\cite{koe92} as well as predictions from model calculations \cite{kol94}. 
Because these cross sections are measured just above the 
muon threshold energy, it is expected that nuclear effects will 
substantially modify the impulse approximation predictions. MEC corrections is one such effect 
and its impact on the impulse approximation calculation is examined in 
this subsection using the soft--pion dominance formalism.

In contrast to the charged current reaction involving the muon--neutrino, 
it is interesting to note that there is good agreement between model predictions and the 
measured cross section for the exclusive charged current reaction $^{12}C(\nu_e,e^-)N_{g.s.}$ 
involving {\em electron}--neutrinos \cite{bod92}.
From the computational point of view, the main difference between 
electron-- and muon--neutrino induced charged current reactions is
that in the former case, the mass of the outgoing electron is negligible while 
the muon mass has to be explicitly included for the latter case. This
means that in the charged current reaction $^{12}C(\nu_{\mu^-},\mu^-)X$, the term 
proportional to the induced axial form factor, $H_A(k^2)$, will contribute 
to the total cross section. In the present formalism, the induced axial
form factor is given by PCAC in the pion pole dominance
\begin{equation}
H_A(k^2) = G_A(0) \frac{2M}{k^2-m^2_\pi}
\label{eq:INDUCED}
\end{equation}
where $G_A(0) \equiv g_A/2 = 1/2(F+D)$ is the axial coupling constant. However, because of the 
presence of the $H_A(k^2)$ term there is an ambiguity when applying 
current conservation to the vector currents. The prescription given by 
Adler in \cite{adl68} has been used here to insure vector current 
conservation and details are given in Appendix~B.

Figure~2 shows the total cross section for the inclusive charged current 
process $^{12}C(\nu_{\mu^-}, \mu^-)X$ for several nuclear densities 
obtained by folding the LSND neutrino energy distribution 
\cite{alb94,LAMPF}. 
As in the electron scattering case, the RFG model without binding energy 
corrections is used to model the nucleus and only $1p1h$ final states are 
considered when taking matrix elements of two--body operators since the 
phase space for $2p2h$ final states should be suppressed due to the 
rather low energy neutrino beam \cite{LAMPF}. 
However, because the flux--weighted average of the neutrino beam energy 
is $<E_\nu>$ = 180 MeV, the three momentum transfers are twice smaller 
than in the electron scattering reaction discussed above and consequently, 
the effect of Pauli blocking becomes more important in this charged current 
reaction. In the RFG model of the nucleus an increase in nuclear density 
is equivalent to an increase in the Fermi momentum. Thus, with a constant 
neutrino beam energy the net effect of Pauli blocking 
is the decrease of the total cross--section {\em per nucleon} with increasing 
nuclear density as shown in Figure~2. In the impulse approximation, the 
total cross section decreases from 24 to 13 $(\times 10^{-40} {\rm 
cm}^2)$ as the Fermi momentum is varied from $k_F$ = 220 to 300 MeV.
As shown in the figure, the inclusion of soft--pion MEC corrections reduces the 
impulse approximation results by 5 to 10\% as the Fermi momentum is increased 
from 220 to 300 MeV. For $k_F$ = 225 MeV, 
which is the usual value used for $^{12}C$, the total cross section is 
reduced from 24.1 to 22.7 $(\times 10^{-40} {\rm cm}^2)$. This reduction is 
not enough to explain the recently measured value reported by the LSND 
collaboration. 

In Figure~3a, the $^{12}C(\nu_{\mu^-}, \mu^-)X$ cross section is shown as a 
function of neutrino energy for $k_F$ = 225 MeV including Coulomb 
corrections for the outgoing muon. Corrections from two--body currents to 
the impulse approximation are very small and difficult to observe. Note 
that the muon production threshold is found to be around 107 MeV compared to 
the experimental threshold of 124 MeV. This is because the present RFG 
calculation does not include any binding energy corrections which will 
ruin the vector current conservation. Corresponding calculation using  
non--relativistic continuum RPA formalism \cite{kol94} obtained a smaller 
cross section which increases more slowly with the neutrino energy. 
The effect of soft--pion MEC is more evident when the differential 
$^{12}C(\nu_{\mu^-}, \mu^-)X$ cross section folded with the experimental
neutrino energy distribution is plotted against the muon 
kinetic energy, $E_\mu$, as shown in Figure~3b. Corrections from two--body currents 
are largest near the cross section peak around $E_\mu$ = 23 MeV where the 
impulse approximation result of $43.7 \times 10^{-42}\;\;{\rm cm}^2/{\rm 
MeV}$ is reduced to $40.1 \times 10^{-42}\;\;{\rm cm}^2/{\rm MeV}$. However, these 
corrections become less important with increasing muon kinetic energy and 
for $E_\mu >$ 60 MeV there is hardly any change from the impulse approximation 
result. The results shown in Figure~3b are qualitatively consistent with non--relativistic 
Fermi Gas and continuum RPA results shown in \cite{kol94} although 
various model dependent assumptions, such as a different neutrino energy 
distribution, prevents from a direct quantitative comparison between 
these results. Thus, although the two--body soft--pion MEC corrections 
help to reduce the impulse approximation prediction of the exclusive 
charged current reaction $^{12}C(\nu_{\mu^-}, \mu^-)X$ towards the observed value, 
this reduction is not enough and additional nuclear structure effects are 
required to further lower the 
total cross section for this exclusive process \cite{kim94,kim95}. These 
additional effects will be incorporated in a future work \cite{udi95}.

\subsection{Quasi--Elastic Neutral Current Neutrino Scattering}
The final application of the soft--pion MEC correction is in the 
quasi--elastic neutral current neutrino--nucleus scattering where 
the one--body neutral current of the constituent nucleon is given by Eq.~(\ref{eq:ZZERO}).
This particular linear combination of the general $SU(3)$ current of 
Eq.~(\ref{eq:GENCUR}) is interesting since, unlike the electromagnetic 
and charged currents, it includes the $SU(3)$ vector and axial {\em singlet} currents leading 
to a possible strange quark contribution to the neutrino--nucleus scattering 
reaction. Also, in contrast to the charged current reactions, the term 
proportional to the induced axial form factor $H_A$ does not contribute 
to the neutral current neutrino--nucleus cross section since the leptons 
involved in the scattering process are massless.
In Eq.~(\ref{eq:ZZERO}), the strange quark electric form factor $F_1^s$ is assumed to 
be vanishingly small for small values of virtuality that 
are of interest here, and therefore only magnetic, $F_2^s$ and axial 
$G_A^s$ strange quark form factors 
of the nucleon are used to describe the contributions from the strange sea quarks 
in neutral current neutrino--nucleus scattering. In the following 
discussion these two strange quark form factors are treated as input parameters 
and no attempt has been made to determine their values from model calculations. 
As in the preceeding discussions on electromagnetic and charged current reactions, 
all the calculations are performed using the RFG model without binding energy corrections.

In Figure~4, differential cross sections for the $^{12}C(\nu, \nu' p)$ 
reaction, $d\sigma/dE_F$, where $E_F$ is the total energy of the ejectile, are shown 
in the quasi--elastic region as functions of the ejected 
proton's kinetic energy, $T_F$, for several values of Fermi momenta.
The incident neutrino energy is assumed to be 200 MeV and the 
values used for the strange magnetic and axial quark form factors are $F_2^s = -0.21$ 
and $G_A^s = -0.19$, respectively. The long dashed curves are the impulse approximation 
results while the solid curves have been obtained with the soft--pion MEC  
corrections, and as in the charged current case, only $1p1h$ final states 
have been taken into account. In the impulse approximation, the magnitude of the 
peak decreases from about 70 to 53 $(\times 10^{-42}\; {\rm cm}^2/{\rm MeV})$ as $k_F$ is 
varied from 200 to 350 MeV. However, this decrease is accompanied by a 
redistribution of the strength to higher values of $T_F$ in such a way 
that the area under the differential cross section remains approximately a constant 
since it is roughly proportional to the number of nucleons which is fixed. 
Note that because the incident neutrino energy is constant and small, the position of the 
maximum of the differential cross section is approximately the same at around 
$T_F$=30 MeV despite the decrease in the peak magnitude.
This means that the effect of Pauli blocking becomes very important when computing the {\em 
total} cross section obtained by integrating the differential cross section
$d\sigma/dE_F$ for $E_F>E_{Fermi}$, thereby cutting off the contributions 
around the peak magnitude.
This total cross section decreases noticeably with increasing $k_F$, as it was the
case for the total charged current cross section shown in Figure~2.

It is evident from the figures that the soft--pion MEC 
effects are sensitive to the nuclear density in this model 
calculation. For $k_F$ around 200 MeV, there is very little difference between the impulse 
approximation and the MEC corrected results 
due to an important cancellation to be discussed below. This 
cancellation becomes less complete with increasing $k_F$ and the two--body 
current effects become stronger and decreases the impulse approximation 
results with increasing nuclear density as shown in the figures. This 
effect of the soft--pion MEC on the impulse approximation persists when the strange quark 
form factors are assumed to vanish ($F_2^s$ = $G_A^s$ = 0) in the model 
calculation. However, without any strange quark form factors there is a large reduction in the differential 
cross section for protons as shown in Figure~5. In this figure $k_F$ is set to 225 MeV and
there is little correction coming from the exchange 
currents in the soft--pion approximation for this value of Fermi momentum. 
Note that the magnitude of the quasi--elastic peak is reduced from about 70 to 50 
$\times 10^{-42}\;{\rm cm^2/MeV}$ when the strange quark magnetic and axial 
form factors are varied from $F_2^s = -0.21$ to $F_2^s$ = 0 and $G_A^s = -0.19$ to 
$G_A^s$ = 0, respectively. For the neutral current induced {\em neutron} 
knockout reaction, $^{12}C(\nu, \nu' n)$, a finite strange quark form factor decreases the 
differential cross section relative to that obtained without any strange 
quark distribution for nucleons \cite{gar92,hor93}. 

It is instructive to understand why the soft--pion exchange current 
corrections are small for neutral current scattering for nuclear densities 
corresponding to $k_F \approx$ 200 MeV. As shown in Appendix~A 
Eq.~(\ref{eq:CRRFGM}), the differential cross section {\em per nucleon} 
for neutral current scattering in the RFG model is given by a linear 
combination of three structure functions $W_L$, $W_T$ and $W_{TT}$ as
\begin{eqnarray}
\left( \frac{d\sigma^{Z^0}}{dE_F} \right)_{\rm RFGM} &= &
\left( \frac{3\pi}{4 k_F^3} \right) \int_{0}^{k_F}\!\! d\epsilon_f 
d(\cos \theta)\;\frac{\sigma^{Z^0}}{|\vec{k\;}|}
\Bigl[ \omega_L W_L+\omega_T W_T+ \omega_{TT'} W_{TT'} \Bigr]  
\label{eq:CRZZERO}
\end{eqnarray}
Here $\omega_L$, $\omega_T$ and $\omega_{TT'}$ are the 
kinematical coefficients corresponding to the longitudinal $W_L$, 
transverse $W_T$ and transverse--transverse $W_{TT'}$ structure 
functions, respectively. 
In Figure~6 the individual contributions from these three structure functions to the 
differential cross section are shown using the same parameters as in Figure~5.
Note that all three contributions to $d\sigma/dE_F$ are positive, 
the dominant one coming from the transverse 
term $\omega_TW_T$ followed by the transverse--transverse $\omega_{TT'} 
W_{TT'}$ and the longitudinal $\omega_L W_L$ terms, the latter 
contributing negligiblely to the differential cross section. Note that the 
soft--pion MEC correction increases the $\omega_{TT'} W_{TT'}$ term and 
decreases the $\omega_T W_T$ term relative to their impulse approximation 
results leading to a cancellation of MEC effects for this particular 
value of Fermi momentum. This cancellation between the 
transverse and transverse-transverse contributions becomes less and less 
complete as the nuclear density is increased and the effect of two--body 
correction becomes bigger for higher densities as shown in Figure~4. 
Eq.~(\ref{eq:CRZZERO}) can also be used to understand the behaviour of the
differential cross section for {\em anti--neutrino} scattering relative 
to the neutrino induced nucleon knockout rates as follows.

Figure~7 shows the $d\sigma/dE_F$ for the neutral current anti--neutrino scattering reaction 
$^{12}C(\bar{\nu}, \bar{\nu}'p)$ and the corresponding neutrino reaction 
both of which are obtained again by using the same input parameters as in Figure~5. 
In the impulse approximation, the magnitude of the quasi--elastic peak for  
anti--neutrino scattering is approximately half that of the neutrino 
scattering, the former value being 38 $\times 10^{-42}\;{\rm cm^2/MeV}$ compared to 
70 $\times 10^{-42}\;{\rm cm^2/MeV}$ for the latter. Also, the position of the peak is shifted 
towards a lower value of $T_F$ by about 10 MeV relative to the neutrino scattering case.
The effect of soft--pion MEC corrections to the impulse approximation 
result for anti--neutrino scattering is noticeably larger than the 
corresponding correction for neutrino scattering for   
nuclear densities corresponding to $k_F$ = 225 MeV as shown in the figure.  
For the $^{12}C(\bar{\nu}, \bar{\nu}'p)$ reaction, the impulse 
approximation result for the magnitude of the quasi--elastic peak
is reduced from 38 to 30 $\times 10^{-42}\;\;{\rm cm^2/MeV}$, while 
there is no such noticeable reduction for the neutrino 
scattering. Thus, not only the anti--neutrino scattering differential cross section greatly reduced 
relative to neutrino scattering, but the effect of soft--pion MEC on 
the impulse approximation result is much bigger reducing 
further the $d\sigma/dE_F$ by about 15\%.

This difference in the neutrino and anti--neutrino 
differential cross sections may be understood by once 
again examining Figure~6 and noting the definitions of the structure functions $W_i$ and their 
corresponding kinematical coefficients $\omega_i$ given in 
Eqs.~(\ref{eq:WL}) to (\ref{eq:wltprime}) of Appendix~A.
The only difference between the longitudinal, transverse and 
transverse--transverse structure functions and kinematical coefficients 
entering in Eq.~(\ref{eq:CRZZERO}) for the neutrino and anti--neutrino 
scattering is in the sign of $\omega_{TT'}$ given by Eq.~(\ref{eq:wttprime}). 
Therefore, the graph corresponding to Figure~6 for anti--neutrino 
scattering will have a negative transverse--transverse contribution
which partially cancels the dominant transverse contribution yielding the 
small differential cross section as shown in Figure~7.
In addition, because the $\omega_{TT'} W_{TT'}$ term is now negative, the 
MEC correction to this term will further decrease the correction to the 
transverse term resulting in a non--negligible two--body correction to 
the differential cross section not observed in the neutrino 
scattering case. Hence although the differential cross section for the 
anti--neutrino reaction $^{12}C(\bar{\nu}, \bar{\nu}'p)$ is about half that of 
the corresponding neutrino scattering reaction, its impulse 
approximation result is subject to non--negligible soft--pion MEC 
corrections.

In the Introduction it was mentioned that the ratio of 
proton--to--neutron yields in quasi--elastic neutral current scattering, 
$R(p/n)$, is currently being measured in the LSND experiment to extract 
the value of the strange quark axial form factor, $G_A^s$. In this 
experiment the probing neutrinos and anti--neutrinos are incident on a 
tank of mineral water composed of hydrogen and carbon molecules 
\cite{gar92}. The amount of kinetic energy carried by the struck nucleon 
is measured by the detectors surrounding the tank of mineral water 
and $R(p/n)$ is determined as a function of the detected ejectile's 
kinetic energy, $T_F$. The proton--to--neutron ratio is then integrated 
over $T_F$ and compared with model predictions which gives the integrated 
$R(p/n)$ as a function of $G_A^s$ for several values of the strange quark 
magnetic form factor as shown in Figure~8. However, in practice there is an 
experimental cutoff in the range of $T_F$, given by 50 MeV $< T_F <$ 120 MeV,
in order to make sure that the detected protons are knocked out from the carbon 
nucleus. The RFG model calculation of the ratio of integrated proton--to--neutron yields 
of Figure~8 is qualitatively similar to the ratios predicted by a non--relativistic 
 RPA calculation \cite{gar92} and shows only about a few percent change from the 
impulse approximation results in the ratio for the neutrino scattering while this change 
is about 10\% for anti--neutrinos. The latter change originates in the 
non--negligible two--body correction to the impulse approximation in the 
anti--neutrino scattering reaction explained in the preceeding paragraph.
In both cases, the two--body 
corrections tend to decrease the integrated ratio and this effect increases as 
$G_A^s$ becomes less and less negative for a given value of $F_2^s$. For 
example, if $G_A^s \approx -0.1$, the soft--pion MEC corrections are negligible for 
the neutrino scattering $R(p/n)$ for all physically relevant values of 
$F_2^s$ but must be taken into account in analyzing the anti--neutrino 
scattering ratio. Finally, final state interaction effects should be examined 
within a relativistic framework, although it is known from a non-relativistic
calculation \cite{gar93} that the effects on the integrated ratios are almost negligible.

\section{Discussion}
In this work relativistic two--body MEC corrections have been applied simultaneously  
to quasi--elastic electromagnetic and neutrino scattering reactions using a single unified 
formalism obtained by generalizing the method developed by Chemtob and Rho \cite{che71}. 
The basic strategy employed here is to use the  chiral filtering hypothesis \cite{kub78,rho81}  
which assumes that the two--body MEC correction in nuclei is dominated by the exchange of 
a single pion whose production amplitude is evaluated in 
the soft--pion limit. By assuming PCAC and using the $SU(3) \otimes SU(3)$ current 
algebra this soft--pion dominance approximation allows one to express the pion production 
amplitudes in terms of $SU(3)$ vector and axial currents interacting with 
the nucleon. Thus the resulting MEC operators are expressed in terms of currents used in the impulse 
approximation without any references to model dependent $N-N$ interaction.
These soft--pion MEC operators arise from the longest range ({\it i.e.} 
the S--wave) part of the $N-N$ interaction and PCAC is used to relate the 
soft--pion predictions to experiment. This is the reason why it is 
necessary to take the spatial part of the pion four momentum to zero 
first when taking the $q_\mu \rightarrow 0$ limit.
The generalized version of the method of Chemtob and Rho used in this work 
is so far the most economical way to estimate exchange current corrections to low and intermediate 
energy neutrino--nucleus scattering since it treats all the $SU(3)$ vector and axial currents 
in Eq.~(\ref{eq:GENCUR}) on the same footing, and is a natural way to accommodate the strange 
quark vector and axial currents of the nucleon allowed by the Standard Model. 

Previously, the non--relativistic and $SU(2)$ version of the soft--pion MEC operators 
have been successfully applied to first forbidden $\beta$--decays 
\cite{kub78,war91,war92,war94} and to electromagnetic and neutrino scattering off few nucleon 
systems \cite{ris72,bac88,tat90,che71} including processes involving large momentum 
transfers \cite{hoc73}. This last phenomenological success came as a 
real surprise since the soft--pion MEC operators were expected to be 
applicable only to those processes involving small momentum transfers such 
as in $\beta$--decays. It led to speculations \cite{rho81} 
that the physics in nuclei is dominated by processes dictated by chiral symmetry
and a theoretical justification was proposed \cite{rho91} to explain the 
phenomenological success by extending the work of Weinberg \cite{wei90}
on nuclear forces derived from chiral Lagrangians \cite{rho91}. Motivated 
by these developments, fully relativistic $SU(3)$ soft--pion MEC operators
have been used to estimate two--body corrections to the impulse 
approximation in quasi--elastic electromagnetic and neutrino scattering 
reactions assuming a finite magnetic and axial strange quark form factors of the nucleon. 
In order to clearly isolated the effect of 
two--body corrections in many body systems, a simple free RFG model of the target nucleus 
was used without any other medium dependent corrections.

The inclusive $(e, e')$ longitudinal and transverse response functions in the quasi--elastic region 
obtained using MEC corrections show a small reduction of the impulse 
approximation result for the longitudinal channel while a substantial 
increase was observed for the transverse response. The former result is in
qualitative agreement with a previous work using non--relativistic kinematics \cite{koh81}, 
while the relativistic MEC corrections tend to increase the 
response function in the transverse channel in contradiction to the 
results of \cite{koh81} but in agreement with another 
relativistic calculation \cite{dek91,dek93}. It should be stressed that 
in almost all calculations of MEC corrections in electron scattering 
reactions, such as in \cite{dek91,koh81}, the one--pion exchange 
operators are of order $g_{\pi NN}^2$ whereas the MEC operators used in 
this work is valid to all orders in $g_{\pi NN}$ at one of the $\pi NN$ 
verticies since the soft--pion 
production amplitude is evaluated non--perturbatively \cite{adl68}.
Thus it is difficult, if not impossible, to make any quantitative 
comparisons with other works dealing with MEC corrections in electron 
scattering especially if additional medium corrections 
are present in a given calculation. However, in order to make a 
more quantitative prediction it is necessary to introduce these
nuclear medium effects, such as density dependent nucleon mass and form 
factors, as well as MEC involving the excitation of the 
$\Delta$ resonance. For these reasons, the most successful application of 
soft--pion MEC correction in electromagnetic interactions is still in the 
electrodisintegration of the deuteron investigated over 20 years ago.

The charged current neutrino--nucleus reaction investigated in the 
present work is the inclusive $^{12}C(\nu_{\mu^-}, \mu^-)X$ reaction
where the total cross section has recently been measured by the LSND 
experiment \cite{alb94}. In the RFG model calculation, the two--body 
soft--pion exchange correction reduces the impulse approximation result 
for the cross section by 5 to 10 \% as the Fermi momentum is increased 
from 200 to 300 MeV. For $k_F$ = 225 MeV, a typical Fermi momentum 
used for $^{12}C$, the total cross section is reduced from 24.1 to 22.7 
$(\times 10^{-40} {\rm cm}^2)$ which is not enough to explain the recently measured value 
reported by the LSND collaboration of (8.3 $\pm$ 0.7 stat. $\pm$ 1.6 
syst.)$\times 10^{-40}{\rm cm}^2$ \cite{alb94}. Recall that the 
previously reported measurement is (15.9 $\pm$ 2.6 stat. $\pm$ 3.7 syst.)$\times 
10^{-39}{\rm cm}^2$ \cite{koe92} which was measured using a different neutrino energy 
distribution than in the LSND experiment. If the LSND measurement is 
correct, then some important density dependent physics is responsible for the 
large reduction of the impulse approximation result, a 
situation reminiscent of the ``missing strength" problem in the 
longitudinal $(e, e')$ response functions of the last decade \cite{orl91}. 

The quasi--elastic neutral current induced nucleon knockout reaction 
$^{12}C(\nu, \nu' N)$ is currently being investigated by the LSND 
collaboration in an effort to extract $G_A^s$ by measuring the ratio of proton--to--neutron yields. 
For the proton knockout reaction, the impulse approximation result for the 
differential cross section is found to be sensitive to the values of 
strange quark form factors of the nucleon, while the two--body soft--pion 
correction becomes more important as the nuclear density is increased.
However, for densities appropriate for $^{12}C$ the MEC corrections to 
the structure functions cancel each other resulting in small changes 
from the impulse approximation results. Because of this cancellation the 
ratio of proton--to--neutron yield for neutrino scattering does not 
suffer from two--body corrections and, at the present level of 
sophistication, the impulse approximation result is sufficient to describe
the ratio $R(p/n)$. The corresponding knockout reaction involving 
anti--neutrinos exhibits a much smaller differential cross section than 
in the neutrino scattering case and a non--negligible MEC correction to the 
impulse approximation result. These differences may be understood by once 
again examining the contributions from different structure functions to 
the differential cross section. In this case, one of the kinematical 
coefficients corresponding to a structure function changes sign relative 
to the neutrino scattering case. As a result, there is a cancellation in 
the impulse approximation results leading to a reduced differential cross 
section but an enhancement in the magnitude of the MEC correction which amounts  
to about 10--15\% of the impulse approximation result depending on 
the value of $G_A^s$. However, in order to make a concrete statement about 
the extraction of the strange quark axial form factor from the measured 
$R(p/n)$, additional nuclear effects, such as final state interactions, need to be 
considered in addition to MEC corrections.

In this paper, motivated by the LSND neutrino scattering experiment, the quasi--elastic 
region in electron and neutrino scattering was chosen to apply the 
soft--pion MEC corrections.
This kinematic regime probably represents the limit of the applicability of 
the soft--pion dominance approximation as presented in this work. 
Description of processes involving larger momentum transfers would 
involve the introduction of the $\Delta$ excitation current, 
pion production processes as well as short ranged meson exchanges. There exists prescriptions to 
extend the present technique to incorporate these processes
as shown in \cite{bac88,tat90} albeit in the non--relativistic 
$SU(2)$ formalism. It would be interesting to extend these prescriptions 
to the present relativistic $SU(3)$ formalism since it can trivially be used to 
estimate MEC corrections to {\em any} linear combination of $SU(3)$ axial and 
vector currents shown in Eq.~(\ref{eq:GENCUR}). For example, the extended 
formalism may be used to investigate parity violating electron 
scattering processes which involves the interference of electromagnetic and
weak neutral currents. Other phenomena of interest to examine are 
neutral current scattering of solar and supernova neutrinos off deuterons 
\cite{bac88} assuming finite strange quark form factors for the nucleons 
and inelastic nuclear transitions which can only take place in the presence finite strange quark 
form factors \cite{suz90}. The latter investigation would involve a 
relativistic finite nucleus neutrino scattering calculation which is the 
subject of the upcoming paper. 
\acknowledgements
YU would like to thank S.~Krewald, C.~Horowitz and K.~Langanke for 
stimulating discussions at their home institutes during the early stages 
of this work. YU was supported by the foundation for Fundamental Research of 
Matter (FOM) and the Dutch National Organization for Scientific Research (NWO). 
YU also acknowledges support from US Department of Energy under grant No. 
DE--FG02--93ER--40762. JMU is carrying out the work as a part of a Community 
training project financed by the European Commission under Contract
ERBCHBICT 920185. 

\appendix
\section{Conventions and Formulae}
In this Appendix conventions and formulae used to compute 
semi--inclusive differential cross--sections for neutral and charged 
current neutrino--nucleus reactions in the RFG model are summarized for completeness. 
Because the technical details for including the $1p1h$ and $2p2h$  MEC contributions to the hadronic 
tensor in the Fermi Gas models are given in detail in \cite{ord81,dek93}, 
only the impulse approximation results are presented here. Although these references deal 
only with electron scattering, generalization to both charged and neutral current neutrino 
scattering is straightforward.
Conventions of Halzen and Martin \cite{HM} are used throughout. 

As usual, it is assumed that the target nucleus consists of 
$A$ nucleons with mass $M_A$ and that a single nucleon is ejected 
from the target due to the interaction with the probe. 
In what follows $M$ is the mass of the free nucleon and the initial and final four 
momenta of the ejectile, $P_I^{\mu} \equiv (E_I, \vec{P}_I)$ and 
$P_F^{\mu} \equiv (E_F, \vec{P}_F)$, are related to the four
momentum transfer to the target $k^{\mu}$ by $P_I^{\mu} = k^{\mu} - P_F^{\mu}$. 
Incoming and outgoing lepton four momenta are denoted by 
$k_i^{\mu} \equiv (\epsilon_i, \vec{k}_i)$ and $k_f^{\mu} \equiv (\epsilon_f, 
\vec{k}_f)$, respectively. A positive energy nucleon with energy $E$ in a plane 
wave state is given by
\begin{equation}
\psi^\sigma_{PW}(\vec{P},\vec{r}\:) = 
\frac{e^{i\vec{P}\cdot\vec{r}}}{\sqrt{2EV}}\:u(\vec{P},\sigma)
\label{eq:PW}
\end{equation}
where the 4--component spinor $u(\vec{P},\sigma)$, normalized to
$u^\dagger u=2E$, is
\begin{equation}
u(\vec{P},\sigma)=\sqrt{E+M} \left(\begin{array}
{@{\hspace{0pt}}c@{\hspace{0pt}}} 
1 \\ 
\displaystyle \frac{\vec{\sigma}\cdot\vec{P}}{E+M}\end{array}\right)\chi^\sigma
\label{eq:SPINOR}
\end{equation}
Here $\chi^\sigma$ is the usual 2-component Pauli spinor with spin 
index $\sigma$ and the plane waves are normalized in a box of volume $V$ such that
\begin{equation}
\int_V d\vec{r}\; (\psi^\sigma_{PW})^\dagger \psi^\sigma_{PW} =1
\end{equation}

The transition matrix element for the neutrino--nucleus scattering 
reaction is given in the usual way in the limit of very heavy vector 
boson masses ({\em i.e.} $m^2_{Z^0,W^\pm} \gg k^2$) as 
\begin{equation}
S_{fi}=i g^2  \int \!\! d^4 x\int \!\! d^4 y
\int \!\! \frac{d^4 k}{(2\pi)^3}\:e^{ik_\alpha (x-y)^\alpha}\: 
j^\mu(\vec{x}\:) J_\mu(\vec{y}\:)
\label{eq:TMX}
\end{equation}
where $j^\mu(\vec{x}\:)$ and $J_\mu(\vec{y}\:)$ 
 are the leptonic and hadronic currents, respectively, and
the coupling constant $g$ is defined as $g^2 \equiv G_F/\sqrt{2}$. 
The spatial parts of the leptonic and hadronic currents in the above transition matrix are
\begin{eqnarray}
j^\mu(\vec{x}\:)  & = & \frac{1}{2V}\frac{e^{i(\vec{k}_i-\vec{k}_f)\cdot \vec{x}}}
{(\epsilon_i\epsilon_f)^{1/2}}\: 
\bar{u}(\vec{k}_f,\sigma_f)   \gamma_\mu (1 \mp \gamma_5) 
u(\vec{k}_i,\sigma_i) 
\label{eq:LEPCUR} \\
J^\mu(\vec{y}\:) & = & \frac{1}{2V}\frac{e^{i(\vec{P}_I-\vec{P}_F) \cdot \vec{y}}}
{(E_I E_F)^{1/2}}\: 
\bar{u}(\vec{P}_F,\sigma_F) \hat{J}^\mu(k) u(\vec{P}_I,\sigma_I)  
\label{eq:HADCUR}
\end{eqnarray}
The minus and plus signs in Eq.~(\ref{eq:LEPCUR}) correspond to neutrino 
and anti--neutrino 
scattering, respectively. In the impulse approximation, the hadronic 
current operator $\hat{J}^\mu(k)$ is given by either Eq.~(\ref{eq:ZZERO}) for 
the neutral current or by Eq.~(\ref{eq:JPM}) for the charged current, while the 
corresponding two--body MEC operators are discussed in Appendix~B.
The resulting cross section for {\em free} nucleons is given by
\begin{equation}
\left( d\sigma^{Z^0/W^\pm} \right)_{\rm Free} = \; 
\delta^{(4)}(k_i^\mu-k_f^\mu+P_I^\mu-P_F^\mu)\;
\sigma^{Z^0/W^\pm} \frac{1}{4\epsilon_f^2 E_I E_F}\: 
\omega_{\mu\nu}W^{\mu\nu}\: d^3\vec{P}_F d^3\vec{k}_f
\label{eq:CROSEC}
\end{equation}
where $\sigma^{Z^0}$ and $\sigma^{W^\pm}$ 
assume the following form for neutral and charged current processes 
\begin{eqnarray}
\sigma^{Z^0} & = &16\: \epsilon_f^2 \cos^2( \theta/2) 
\left[\frac{g^2}{4\pi}\right]^2 \\
\sigma^{W^\pm} & = & 16 k_f^2 \left[ \frac{g^2}{4\pi} \right]^2  
\end{eqnarray}
with $\theta$ being the lepton polar scattering angle. The quantity $\omega_{\mu\nu} W^{\mu\nu}$ 
appearing in Eq~(\ref{eq:CROSEC}) may be expressed as a linear combination of various nuclear 
structure functions $W_i$ as
\begin{equation}
\omega_{\mu\nu} W^{\mu\nu} = \omega_L W_L+\omega_T W_T+ \omega_{TT} W_{TT} +
\omega_{LT} W_{LT} + \omega_{LT'} W_{LT'}+\omega_{TT'} W_{TT'}
\label{eq:STRUCTURE}
\end{equation}

These structure functions are most conveniently expressed in a coordinate 
system defined by the following orthogonal unit vectors 
\begin{eqnarray}
\hat{z} & \equiv & \hat{k} \\ 
\hat{n}_\perp & \equiv &\frac{\vec{k}\times \vec{P}_F}{|\vec{k}\times \vec{P}_F|}
\label{perp} \\
\hat{n}_\parallel & \equiv &\frac{\hat{n}_\perp \times \vec{k}}{|\hat{n}_\perp 
\times \vec{k}\:|}
\label{parp}
\end{eqnarray}
In this coordinate system the spatial part of the hadronic current 
$\vec{J}$, where $J^\mu \equiv (\rho, \vec{J}\;)$, may be written as
\begin{equation}
\vec{J} =J_\parallel \hat{n}_\parallel+J_\perp \hat{n}_\perp+J_k \hat{k}
\end{equation} 
Then the structure functions for the semi--inclusive neutral current 
$(\nu,\nu'N)$ reaction are
\begin{eqnarray}
W_L &=& |\rho|^2 +\frac{\omega^2}{|\vec{k\:}|^2} |J_k|^2 - 
\frac{\omega}{|\vec{k}\:|}\; 2 Re\:(\rho^{*}J_k) 
\label{eq:WL} \\
\omega_T W_T &=& \omega_\perp W_\perp + \omega_\parallel W_\parallel \\
W_\perp &=& |J_\perp|^2 \\
W_\parallel &=& |J_\parallel|^2 \\
W_{TT} &=& \sin(2\phi_F) Re\:(J_\perp J^\dagger_\parallel) \\
W_{LT} &=& Re\left[ \Bigl( \rho-\frac{\omega}{|\vec{k}\:|}J_k \Bigr)
\Bigl( -\sin \phi_F J_\perp^\dagger+\cos \phi_F J_\parallel^\dagger \Bigr) \right] \\
W_{TT'} &=& Im\:(J_\parallel J_\perp^\dagger) \\
W_{LT'} &=& Im \left[ \Bigl( \rho-\frac{\omega}{|\vec{k}\:|}J_k \Bigr)
\Bigl( \sin \phi_F J_\parallel^\dagger+\cos \phi_F J_\perp^\dagger \Bigr) \right]
\end{eqnarray}
where $\phi_F$ is the azimuthal angle of the ejected nucleon. The 
kinematical coefficients $\omega_i$ corresponding to these structure 
functions are as  follows
\begin{eqnarray}
\omega_L &=& 1 \\
\omega_\perp &=& \tan^2 (\theta/2) - \Bigl[ 1+\cos (2\phi_F) \Bigr] 
\left( \frac{k_\mu^2}{2|\vec{k\:}|^2} \right) \\
\omega_\parallel &=& \tan^2(\theta/2) - \Bigl[1-\cos(2 \phi_F) \Bigr]
\left( \frac{k_\mu^2}{2|\vec{k\:}|^2} \right) \\
\omega_{TT} &=& \frac{k_\mu^2}{|\vec{k\:}|^2} \\
\omega_{LT} &=& -2 \sqrt{\tan(\theta/2)- \frac{k_\mu^2}{|\vec{k\:}|^2}}
 \\
\omega_{TT'} &=& \mp 2 
\tan(\theta/2)\sqrt{\tan(\theta/2) - \frac{k_\mu^2}{|\vec{k\:}|^2}}
\label{eq:wttprime} \\
\omega_{LT'} &=& \pm 2 \tan(\theta/2)
\label{eq:wltprime}
\end{eqnarray}
In the expressions for the coefficients $\omega_{TT'}$ and $\omega_{LT'}$, the upper and 
lower signs correspond to neutrino and anti--neutrino scatterings, respectively. 
For the charged current reaction of type $(\nu_{l},l N)$ with lepton 
mass $m_l$, each term in the product $\omega_{\mu\nu} W^{\mu\nu}$, 
Eq.~(\ref{eq:STRUCTURE}), may be written as
\begin{eqnarray}
\omega_L W_L &=&\frac{1}{4\epsilon_i k_f} \Biggl\{ \:\Biggl[ (\epsilon_i+\epsilon_f)^2-|\vec{k\:}|^2-m_l^2 
\Biggr] 
|\rho|^2 \Biggr. \nonumber \\
                       & & \;\;\;\; +\:\left[ \frac{(\epsilon_i^2-k_f^2)^2}{|\vec{k\:}|^2} - 
\omega^2+ m_l^2 \right] |J_k|^2 \nonumber \\
                       & & \Biggl. \;\;\;\;\;\;\;\; -\:\left[ \frac{2(\epsilon_i+e_f)(\epsilon_i^2-k_f^2)}{|\vec{k}\:|} 
- 2\omega |\vec{k}\:| \right] Re\:(\rho^* J_k) \Biggr\} \\
\omega_T W_T &=&   \left\{
\frac{\epsilon_i k_f \sin^2\theta}{2 |\vec{k\:}|^2} \cos(2\phi_F)
\Bigl( |J_\parallel|^2-|J_\perp|^2 \Bigr)\right. \nonumber \\ 
                       &  &  \;\;\;\; +\left.\left[\frac{\epsilon_i k_f \sin^2\theta}{2|\vec{k\:}|^2}-\frac{1}{2}
\left(\frac{-\epsilon_f}{k_f}+\cos\theta \right) \right] 
\Bigl( |J_\parallel|^2+|J_\perp|^2 \Bigr) \right\} \\
\omega_{LT} W_{LT} &=& \frac{ \sin \theta}{|\vec{k}\:|}
\Bigl( \epsilon_i+\epsilon_f \Bigr) \nonumber \\
                       & &  \;\;\;\; \times Re\left[ \left(\frac{\epsilon_i^2-k_f^2}{|\vec{k}\:|(\epsilon_i+
\epsilon_f)} J_k^*-\rho^* \right) \Bigl(J_\parallel \cos\phi_F - J_\perp 
\sin \phi_F \Bigr) \right] \\
\omega_{TT} W_{TT} &=& -
\frac{\epsilon_i k_f\sin^2\theta}{|\vec{k\:}|^2}
\sin (2\phi_F) Re\:(J_\perp J_\parallel^*) \\
\omega_{LT'} W_{LT'} &=& - \sin \theta\;
Im\left[\left(\rho^* -\frac{\omega}{|\vec{k}\:|} J_k^*\right)
\Bigl( J_\parallel \sin\phi_F + J_\perp \cos\phi_F \Bigr) \right] \\
\omega_{TT'} W_{TT'} &=& - \frac{1}{|\vec{k}\:|}
\left(\frac{\epsilon_i \epsilon_f}{k_f}+k_f- 
\bigl(\epsilon_i+\epsilon_f\bigr)\cos \theta \right) Im\:(J_\parallel J_\perp^*) 
\end{eqnarray}

In both the charged and neutral current processes the 4-momentum of the 
scattered lepton  is not observed and it is thus necessary to integrate over 
the unobserved final lepton 3-momentum $\vec{k}_f$ in addition to the angular variables
of the ejected nucleon. This leads to the following differential cross section 
for {\em free} nucleons
\begin{equation}
\left( \frac{d\sigma^{Z^0/W^\pm}}{dE_F} \right)_{\rm Free} 
=\frac{\sigma^{Z^0/W^\pm}}
{4\epsilon_f^2}\frac{2\pi}{|\vec{k}\:|}
\Bigl( \omega_L W_L+\omega_T W_T+ \omega_{TT'} W_{TT'} \Bigr)
\end{equation}
In the relativistic FGM, non--interacting nucleons are filled up to the Fermi momentum 
$k_F$ and the initial momentum of the struck nucleon have to be 
averaged over the Fermi sphere.
In addition, the initial energy of this nucleon $E_I$ is often reduced 
with respect to its free space value by 
the binding energy, expressed by an input parameter $B$ 
\begin{eqnarray}
E_I & = & \Bigl( |\vec{P_I\:}|^2+M^2 \Bigr)^{1/2}-B \nonumber \\
      & = & \Bigl( |\vec{k}|^2+|\vec{P}_F|^2-2\:\vec{k}\cdot\vec{P}_F+M^2 \Bigr)^{1/2}-B 
\end{eqnarray}
The resulting differential cross sections {\em per nucleon} for neutral and charged current 
processes in the RFGM are
\begin{eqnarray}
\left( \frac{d\sigma^{Z^0/W^\pm}}{dE_F} \right)_{\rm RFGM} &= &
\left( \frac{3\pi}{4 k_F^3} \right) \int_{0}^{k_F}\!\! d\epsilon_f d(\cos \theta)\;
\frac{\sigma^{Z^0/W^\pm}}{|\vec{k\;}|}
\Bigl[ \omega_L W_L+\omega_T W_T+ \omega_{TT'} W_{TT'} \Bigr]  
\label{eq:CRRFGM}
\end{eqnarray}
Integrations over $\epsilon_f$ and $ \cos\theta$ in Eq.~(\ref{eq:CRRFGM}) 
are performed numerically. For further technical details see \cite{jose95}.
\section{Two--Body Exchange Current Operators}
This Appendix presents the general form for the two--body soft--pion MEC operator, 
$J_\mu^a(k; P_{I,1}; P_{I,2}; P_{F,1}; P_{F,2})_{EX}$,
given in Eq.~(\ref{eq:JEX}) where the probing current, $J^a_\mu(k)$, 
may be any one of vector or axial $SU(3)$ currents. Since the matrix element 
for the pion absorption is given by Eq.~(\ref{eq:PINN}), the quantity of 
interest here are the amplitudes for the soft--pion production shown in 
Eq.~(\ref{eq:LSZ}). Therefore, only these amplitudes are
shown explicitly here and the relevant full soft--pion MEC operators for electromagnetic, neutral 
and charged current scattering reactions may be constructed by taking 
the appropriate linear combinations of the operators given in this 
appendix, multiplied by the pion propagator and the $\pi NN$ vertex 
function as shown in Eq.~(\ref{eq:JEX}). 

Using the notation introduced in the text and in Appendix~A, the 
amplitude for soft--pion production induced by the vector singlet current 
$V^0_\mu$ is found to be
\begin{eqnarray}
\lim_{q \rightarrow 0} \langle \pi^a(q) N(P_F) | V^0_\mu(k) | N(P_I) 
\rangle  & = & \nonumber \\ 
&    & 
\!\!\!\!\!\!\!\!\!\!\!\!\!\!\!\!\!\!\!\!\!\!\!\!\!\!\!\!\!\!\!\!\!\!\!\!\!\!\!\!\!\!\!\!\!\!\!\!\!\!\!\!\!\!\!\!\!\!\!\!\!\!\!\!\!\! 
(-i) \bar{u}(\vec{P}_F, \sigma_F) 
\Bigl[ V^0_A \gamma_5\gamma_0\gamma_\mu +
V^0_B \gamma_5\gamma_\mu\gamma_0 + 
V^0_C \gamma_5\gamma_0\Sigma_\mu  \nonumber \\
&  &
\;\;\;\;\;\;\;\;+ V^0_D \gamma_5\Sigma_\mu\gamma_0 +
V^0_E \gamma_5\Sigma_\mu
\Bigr] \lambda^a u(\vec{P}_I, \sigma_I)
\label{eq:MVSING}
\end{eqnarray}
In the above expression $a=1,2,3$ for pion production and the coefficients 
$V^0_J$ multiplying the operators are defined as
\begin{eqnarray}
V^0_A & \equiv & +\sqrt{\frac{2}{3}} F^0_1(k^2) \left( \frac{g_{\pi NN} }{2 E_2} \right)\\
V^0_B & \equiv & +\sqrt{\frac{2}{3}} F^0_1(k^2) \left( \frac{g_{\pi NN} }{2 E_1} \right) \\
V^0_C & \equiv & +\sqrt{\frac{2}{3}} F^0_2(k^2) \left( \frac{g_{\pi NN} }{2 E_2} \right) \\
V^0_D & \equiv & - \sqrt{\frac{2}{3}} F^0_2(k^2) \left( \frac{g_{\pi NN} }{2 E_1} \right) \\
V^0_E & \equiv & +\sqrt{\frac{2}{3}} F^0_2(k^2)\left( \frac{g_{\pi NN} }{M} \right) \\
\end{eqnarray}
with $E_i \equiv \sqrt{|\vec{P_i}\,|^2 + M^2}$. The vector singlet amplitude, 
Eqs.~(\ref{eq:MVSING}) is divergenceless {\em when the nucleons are
assumed to obey the free Dirac equation}, leading to current conservation 
for vector singlet MEC. The corresponding vector octet soft--pion production amplitude is 
\begin{eqnarray}
\lefteqn{\lim_{q \rightarrow 0} \langle \pi^a(q) N(P_F) | V^b_\mu(k) | 
N(P_I) \rangle =} \nonumber \\
&  & 
(+i) \bar{u}(\vec{P}_F,\sigma_F) \left\{ \frac{1}{4} \Bigl[ \lambda^a, \lambda^b \Bigr]_+ \Bigl( 
V^b_A \gamma_5\gamma_0\gamma_\mu +
V^b_B \gamma_5\gamma_\mu\gamma_0 + 
V^b_C \gamma_5\gamma_0\Sigma_\mu + 
V^b_D \gamma_5\Sigma_\mu\gamma_0 +
V^b_E \gamma_5\Sigma_\mu \Bigr) \right.
\nonumber\\
&   & 
\;\;\;\;\;\;\;\;\;\;\;\;\;\;\;\;\;\;\;\;\;\, +\frac{1}{4} \Bigl[ \lambda^a, \lambda^b 
\Bigr]_- \Bigl(
V^b_A \gamma_5\gamma_0\gamma_\mu -
V^b_B \gamma_5\gamma_\mu\gamma_0 + 
V^b_C \gamma_5\gamma_0\Sigma_\mu  -
V^b_D \gamma_5\Sigma_\mu\gamma_0 +
V^b_F \gamma_5\gamma_\mu \nonumber \\
&   &
\;\;\;\;\;\;\;\;\;\;\;\;\;\;\;\;\;\;\;\;\;\;\;\;\;\;\;\;\;\;\;\;\;\;\;\;\;\;\;\;\;\;\;\;\;\;\;\;\;\;\;\;\;\;\;\;\;\;\;\;\;\;\;\;\;\;\;\;\;\;\;\;\;\;\;\;\;\;\;\;\;\;\;\;\;\;\;\;\;\;\;\;
\left. + V^c_0 \gamma_5 k_\mu \Bigr)\right\} 
u(\vec{P}_I,\sigma_I)
\label{eq:MVOCT}
\end{eqnarray}
where the index $b$ runs from 1 to 8. The coefficients $V^b_J$ are 
\begin{eqnarray}
V^b_A & \equiv & +F_1^b(k^2) \left( \frac{g_{\pi NN}}{E_2} \right)\\
V^b_B & \equiv & +F_1^b(k^2) \left( \frac{g_{\pi NN}}{E_1} \right)\\
V^b_C & \equiv & +F_2^b(k^2) \left( \frac{g_{\pi NN}}{E_2} \right)\\
V^b_D & \equiv & -F_2^b(k^2) \left( \frac{g_{\pi NN}}{E_1} \right)\\
V^b_E & \equiv & +F_2^b(k^2) \left( \frac{2 g_{\pi NN}}{M} \right)\\
V^b_F & \equiv & +\left( F_1^b(k^2) + 
\frac{G^b_A(k^2)}{g_A} \right) \left( \frac{2 g_{\pi NN}}{M}\right)\\
V^b_0 & \equiv & \left( \frac{F_\pi^b(k^2)}{k^2 - m_\pi^2} - \frac{H_A^b(k^2)}{g_A} \right)
\left( 2 g_{\pi NN}\right)
\end{eqnarray}
The first term in the coefficient $V^b_0$ yields the ``pion--in--flight'' 
MEC with the pion form factor $F_\pi(k^2)^b$. However, this 
``pion--in--flight'' term is proportional to the four momentum transfer of the 
leptonic probe, $k_\mu$, and thus does not contribute to the neutral current 
total cross section. It also does not contribute to the electron scattering 
total cross section when the usual approximation of neglecting the 
electron mass is made. Terms proportional to $k_\mu$ also do not contribute 
to the $(e, e')$ longitudinal and transverse response functions $R_L$ and 
$R_T$, when they are evaluated by using all the components of the 
electromagnetic four current as done in this paper. The coefficient $V^b_0$ can be made to 
vanish by assuming a point like pion and using the pion pole dominance 
approximation from PCAC, Eq.~(\ref{eq:INDUCED}). For momentum
transfers ranging from $\beta$--decay to the quasi--elastic kinematics
where the soft--pion MEC operators are applicable, this should be a good
approximation.

However, even when $V^b_0$ is made to vanish the vector octet soft--pion 
production amplitude, Eq.~(\ref{eq:MVOCT}), is not divergenceless. The 
problem is the term with coefficient $V^b_F$ which has the structure of 
an axial current. The prescription used by Adler \cite{adl68} to make 
Eq.~(\ref{eq:MVOCT}) divergenceless and thus guarantee the conservation of vector currents 
involves adding appropriate counter terms proportional to $k_\mu$. 
In the present case, this counter term is $V^b_G \gamma_5 k_\mu$ where $V^b_G$ is given by
\begin{equation}
V^b_G  \equiv +\left( \frac{G^b_A(k^2)}{g_A} \right) \left( \frac{4 g_{\pi NN}}{k^2}\right)
\end{equation}
This prescription is applicable to neutral current and electromagnetic reactions 
since terms proportioinal to $k_\mu$ do not contribute to the total cross 
sections. However, it leads to ambiguities for the charged current 
reactions where one of the leptons is massive and terms proportional to 
$k_\mu$ in the hadronic current must explicitly be taken into account in 
the derivation of the reaction cross section. 
Nevertheless, since the generalized soft--pion dominance method of 
Chemtob and Rho ought to be applicable for all types of currents, it is desirable to insure the 
conservation of vector currents in this manner. In the present 
application, the contribution from the counter term $V^b_G \gamma_5 
k_\mu$ to the charged current cross section was found to be numerically 
very small (about 1\% of the total MEC contribution). 

The matrix element of interest for the probing axial singlet current may be written as
\begin{eqnarray}
\lefteqn{\lim_{q \rightarrow 0} \langle \pi^a(q) N(P_F) | A^0_\mu(k) | 
N(P_I) \rangle =} \nonumber \\ 
& & \;\;\;\;\;\;\;(+i) \bar{u}(P_F,\sigma_F) \Bigl[ A^0_A\gamma_0\gamma_\mu + A^0_B \gamma_\mu 
\gamma_0 + A^0_C k_\mu + A^0_D \gamma_0 k_\mu \Bigr] \lambda^a u(P_I,\sigma_I) 
\label{eq:MASING}
\end{eqnarray}
where,
\begin{eqnarray}
A^0_A & \equiv & +\sqrt{\frac{2}{3}} G^0_A(k^2) \left( \frac{g_{\pi NN}}{2 E_2} \right)\\
A^0_B & \equiv & -\sqrt{\frac{2}{3}} G^0_A(k^2) \left( \frac{g_{\pi NN}}{2 E_1} \right)\\ 
A^0_C & \equiv & +\sqrt{\frac{2}{3}} H^0_A(k^2) \left( \frac{g_{\pi NN}}{M}\right)\\ 
A^0_D & \equiv & -\sqrt{\frac{2}{3}} H^0_A(k^2) \left( \frac{g_{\pi NN}}{2}\right)
\left( \frac{1}{E_1} + \frac{1}{E_2} \right)
\end{eqnarray}
Similarly, the corresponding amplitude for the axial octet current is
\begin{eqnarray}
\lefteqn{\lim_{q \rightarrow 0} \langle \pi^a(q) N(P_F) | A^b_\mu(k) | 
N(P_I) \rangle =} \nonumber \\
&  & (+i) \bar{u}(P_F,\sigma_F) \left\{ \frac{1}{4} \Bigl[ \lambda^a, 
\lambda^b \Bigr]_+ 
\Bigl( A^b_A \gamma_0\gamma_\mu + 
A^b_B \gamma_\mu \gamma_0 +
A^b_C \gamma_0 k_\mu + 
A^b_D k_\mu \Bigr)\right. \nonumber\\
&  & \;\;\;\;\;\;\;\;\left. +\frac{1}{4} \Bigl[ \lambda^a, \lambda^b \Bigr]_-
\Bigl( A^b_A \gamma_0\gamma_\mu - 
A^b_B \gamma_\mu\gamma_0 + 
A^b_E \gamma_0 k_\mu + 
A^b_F (p_1 + p_2)_\mu +
A^b_G \gamma_\mu \Bigr) \right\} u(P_I,\sigma_I) 
\label{eq:MAOCT}
\end{eqnarray}
with the following coefficients
\begin{eqnarray}
A^b_A & \equiv & + G^b_A(k^2) \left( \frac{g_{\pi NN}}{E_2} \right)\\
A^b_B & \equiv & - G^b_A(k^2) \left( \frac{g_{\pi NN}}{E_1} \right) \\
A^b_C & \equiv & -H^b_A(k^2) g_{\pi NN} \left( \frac{1}{E_1} + 
\frac{1}{E_2} \right) \\
A^b_D & \equiv & +H^b_A(k^2) \left( \frac{2g_{\pi NN}}{M} \right) \\
A^b_E & \equiv & -H^b_A(k^2) g_{\pi NN} \left( \frac{1}{E_2} - 
\frac{1}{E_1} \right) \\
A^b_F & \equiv & +F^b_2(k^2) \left( \frac{g_{\pi NN}}{g_A M^2} \right) \\
A^b_G & \equiv & -\Bigl( F^b_1(k^2) + F^b_2(k^2) + g_AG^b_A(k^2)  \Bigr)
\left( \frac{2 g_{\pi NN}}{g_A M}\right) 
\end{eqnarray}
Note that it is straightforward to generalize these soft--pion production 
amplitudes to pseudoscalar octet soft--meson production amplitudes where the $SU(3)$ 
index $a$ now takes on values from 1 to 8 and the $\pi NN$ coupling 
constant is replaced by the appropriate meson--nucleon coupling constants.
However, in deriving the above amplitudes the Goldberger--Trieman relation was used 
to simplify the expressions and this algebraic manipulation is not necessarily justifiable for other 
pseudoscalar octet mesons under consideration. 
%
%

%
\vfill\eject
%
%
\centerline{\bf FIGURE CAPTIONS}
\vskip 1cm
FIG~1. \hspace{0.1 cm} a) Longitudinal, $R_L$, and b) transverse, $R_T$, response functions for
the $^{12}C(e, e')$ reaction plotted against the energy transfer $\omega$. 
The three momentum transfer involved in the inclusive reaction is 
$|\vec{k}\;|$ = 400 MeV. A RFG model without any binding energy corrections ($B=0$) and 
with a Fermi momentum of $k_F$ = 225 MeV was used to calculate
the response functions .
Experimental data, taken from \cite{bar83}, are shown for reference but no attempt to fit the data by 
varying $B$ and $k_F$ has been made.
The dashed line represents the impulse approximation results while 
response functions obtained with soft--pion MEC corrections are shown with solid lines. Also 
shown in the figures using the dashed-dotted lines are the individual contributions from soft--pion 
two--body courrents. 
\vskip 0.75cm
FIG~2 \hspace{0.1 cm} $^{12}C(\nu_{\mu^-},\mu^- )X$ total cross section obtained by folding 
the LSND neutrino flux \cite{alb94} versus the Fermi momentum $k_F$. The dashed curve 
is the impulse approximation result while the solid curve is obtained 
with the soft--pion MEC corrections using the RFG model of the nucleus 
with no binding energy correction ($B=0$).  For $k_F$ = 225 MeV, 
which is the usual value used for $^{12}C$, the total cross--section is 
reduced from 24 to 22.7 $(\times 10^{-40} {\rm cm}^2)$ when corrected for two--body 
current effects.
\vskip 0.75cm
FIG.~3 \hspace{0.1 cm} a) $^{12}C(\nu_{\mu^-},\mu^-)X$ total cross section 
as a function of incoming neutrino energy $E_\nu$ taking into account the Coulomb 
correction for the outgoing muon. As in Figure~2, the RFG model is used 
to model the nucleus with $k_F$ = 225 MeV and $B=0$. The dashed line is the impulse 
approximation result while the solid line includes two--body soft--pion MEC effects.
Note that in this particular case there is 
very little difference between the two results. b) Differential cross section 
$d\sigma/dE_\mu$ for the same process obtained with and without the 
soft--pion MEC correction plotted against the outgoing muon kinetic energy 
$E_{\mu}$. The results for this figure is obtained by folding the LSND neutrino energy 
distribution obtained from \cite{alb94}. 
\vskip 0.75cm
FIG~4 \hspace{0.1 cm} $^{12}C(\nu,\nu'p)$ differential cross section versus 
the kinetic energy of the ejected nucleon, $T_F$, for a) $k_F$ = 200 MeV, 
b) $k_F$ = 300 MeV and c) $k_F$ = 350 MeV. The incident neutrino energy is 
assumed to be 200 MeV while the 
values for the strange quark form factors used are $F_2^s = -0.21$ and $G_A^s 
= -0.19$. The long dashed curve is the impulse approximation result while the 
solid curves have been obtained with soft--pion MEC corrections. 
\vskip 0.75cm
FIG.~5 \hspace{0.1 cm} $^{12}C(\nu,\nu'p)$ differential cross section versus 
the kinetic energy of the ejected nucleon, $T_F$, with ($F_2^s = -0.21$ and $G_A^s 
= -0.19$) and without finite strange quark form factors. The incident neutrino energy is 
assumed to be 200 MeV and the long dashed curve is the impulse approximation result while the 
solid curves have been obtained with soft--pion MEC corrections. 
\vskip 0.75cm
FIG.~6 \hspace{0.1 cm} Various structure function contributions to the $^{12}C(\nu,\nu'p)$ 
differential cross section shown in Figure~5. Both the impulse approximation and MEC corrected 
results for the transverse, $\omega_T W_T$, longitudinal, $\omega_L W_L$, and 
transverse--transverse, $\omega_{TT'} W_{TT'}$, contributions are shown explicitly 
using the dashed and solid lines, respectively.
Note that the two--body corrected contributions 
from the transverse and transverse--transverse contributions cancel leading to a small overall 
MEC correction to the impulse approximation. For anti--neutrino 
scattering the transverse--transverse contribution changes sign resulting in a smaller differential cross 
section but in a larger MEC effect compared to neutrino scattering.
\vskip 0.75cm
FIG.~7 \hspace{0.1 cm} $^{12}C(\nu,\nu'p)$ and $^{12}C(\bar{\nu}, \bar{\nu}'p)$ differential 
cross sections versus the kinetic energy of the ejected nucleon, $T_F$. The incident neutrino energy is 
200 MeV and the nucleon is assumed to have strange quark form factors of $F_2^s = -0.21$ and $G_A^s 
= -0.19$. The long dashed curve is the impulse approximation result while the 
solid curves have been obtained with soft--pion MEC corrections. 
\vskip 0.75cm
FIG.~8 \hspace{0.1 cm} a) Ratios of integrated proton--to--neutron quasi--elastic yield 
for the $^{12}C(\nu,\nu'N)$ reaction as functions of $G_A^s$ for two values 
of strange magnetic form factor $F_2^s$.
In each case, the dashed line is the impulse approximation result while 
the solid line has been corrected for meson exchange currents. The incident neutrino 
energy is assumed to be 200 MeV for both cases and the range of integration was chosen to be 
$50 \leq T_F \leq 120$ MeV to simulate the LSND experiment \cite{gar92}.
b) Same as in a) but for anti--neutrino scattering. 
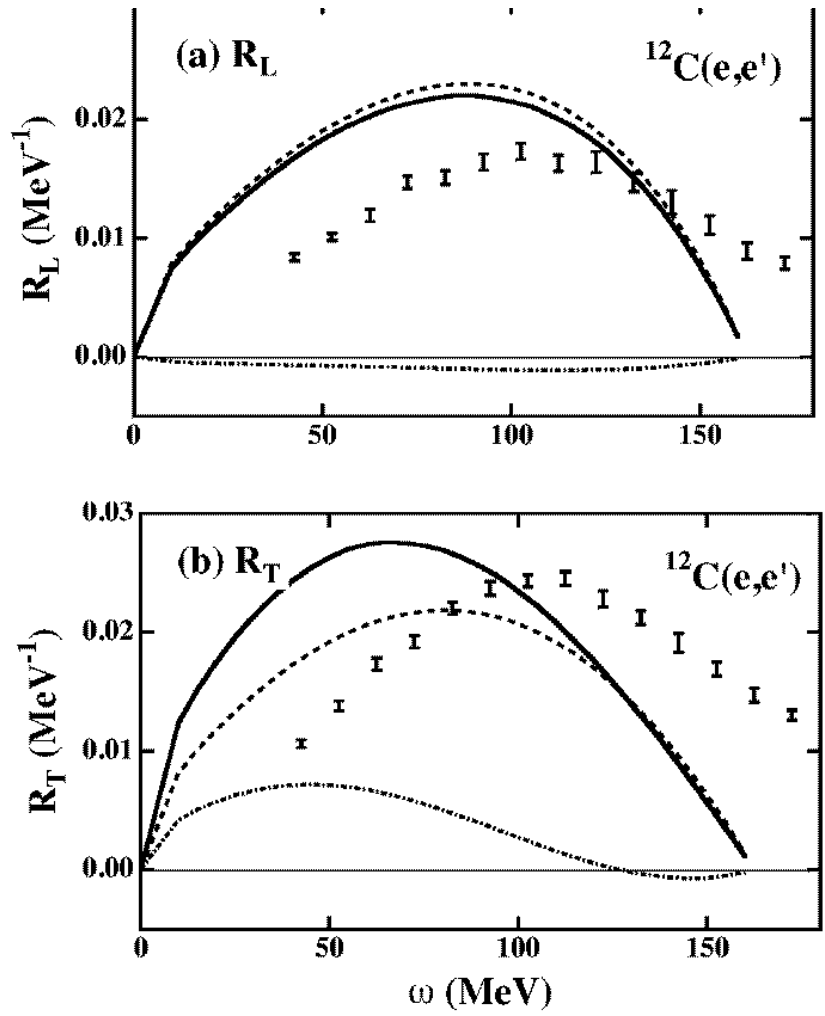
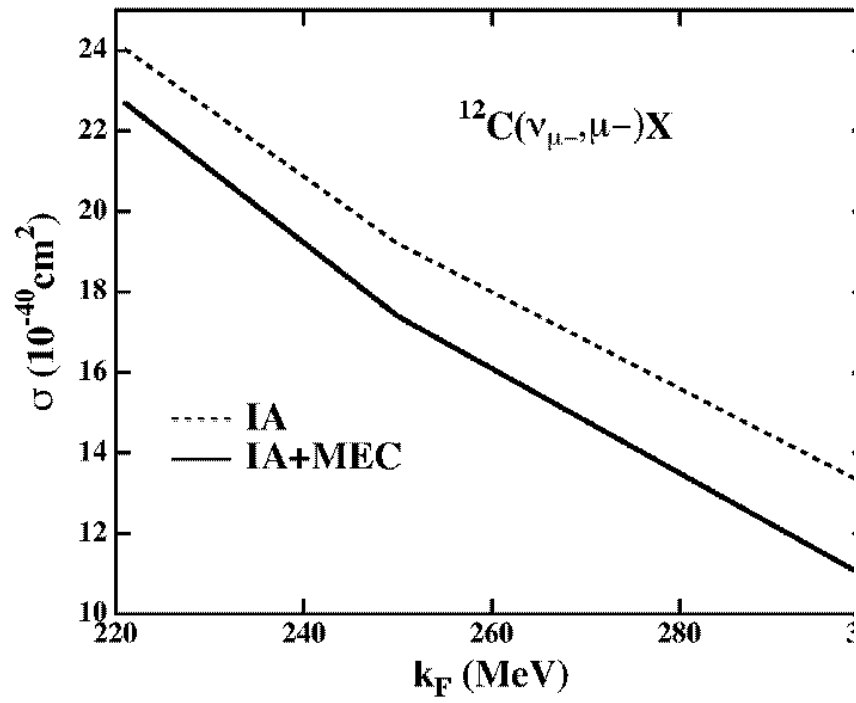
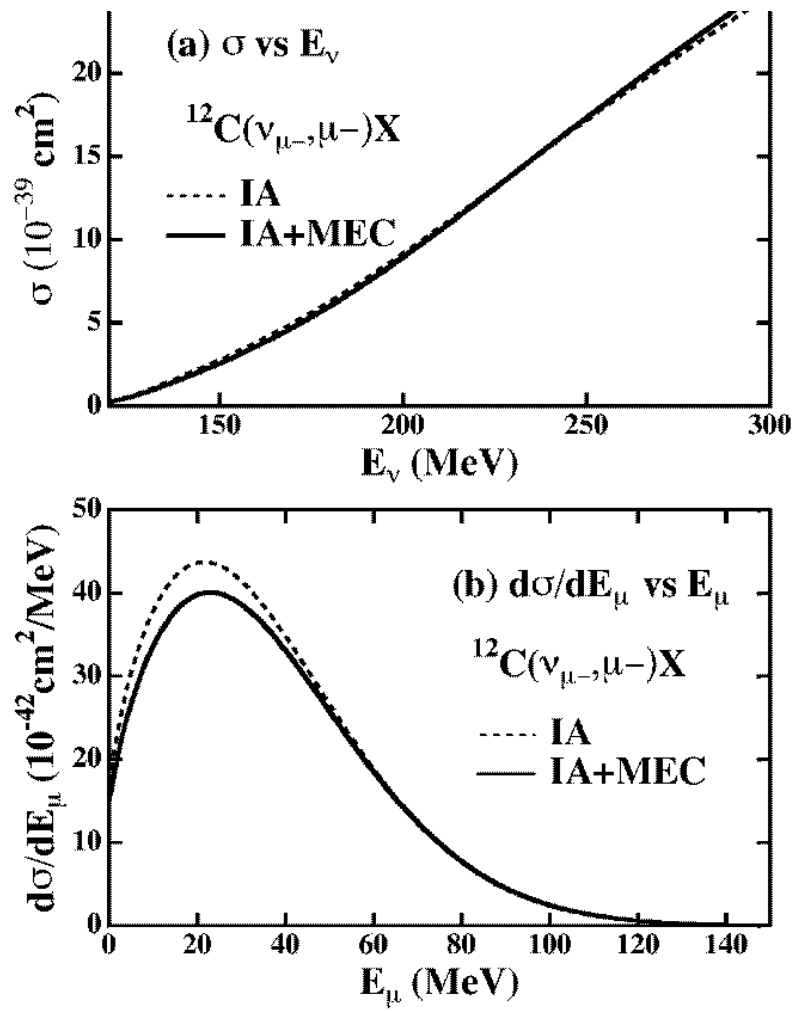
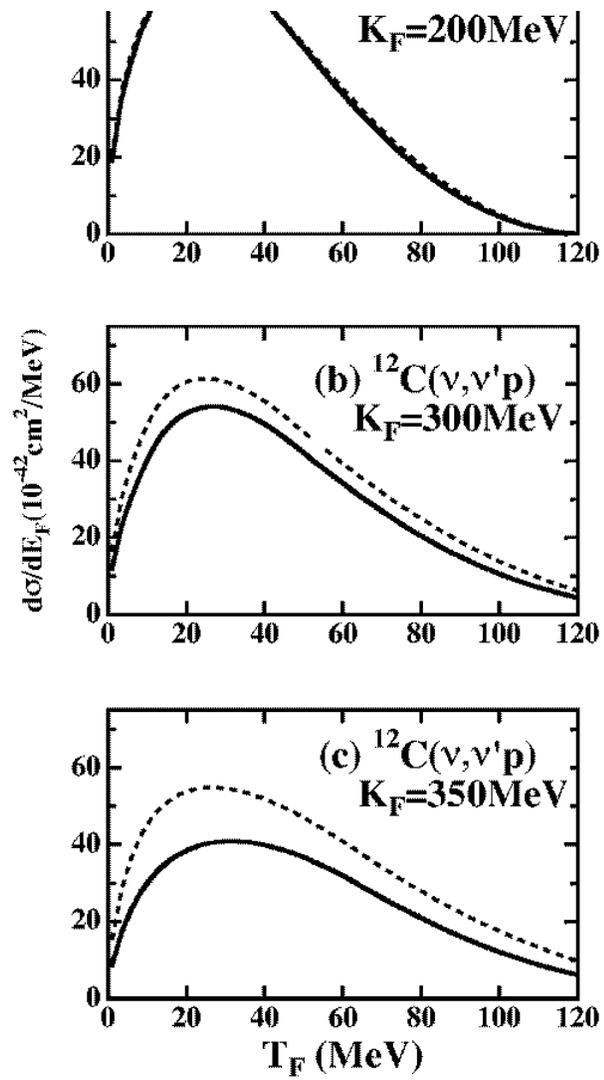
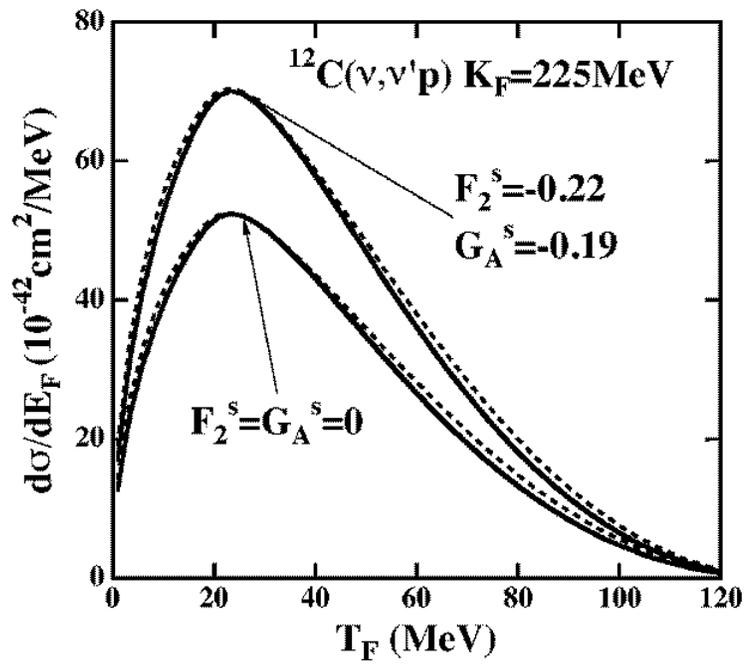
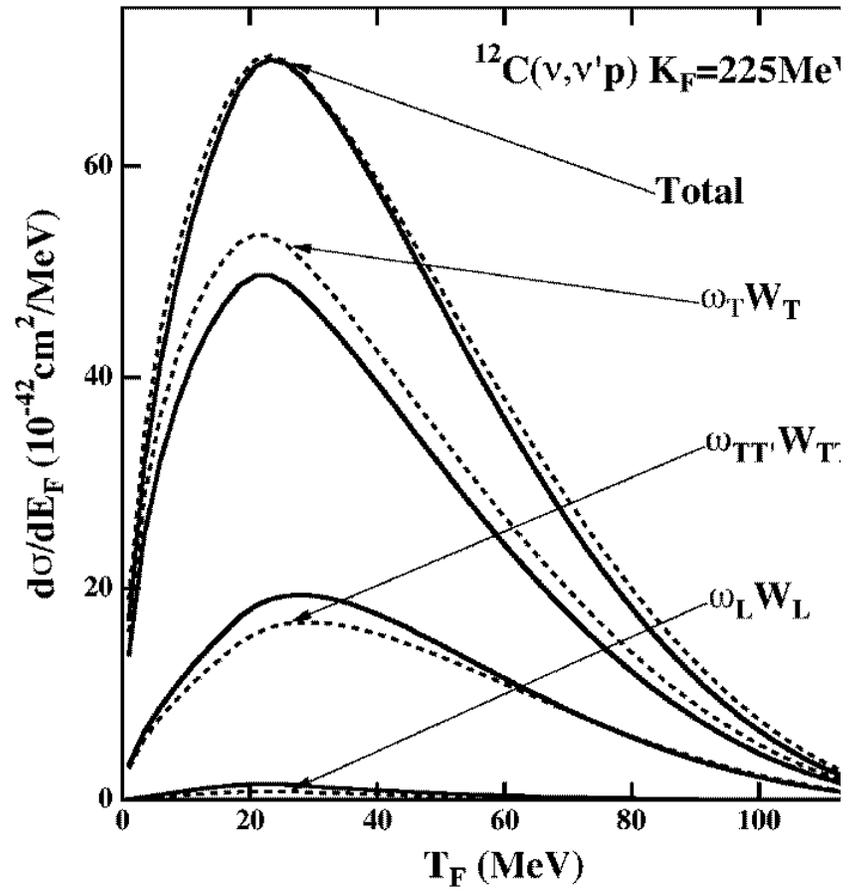
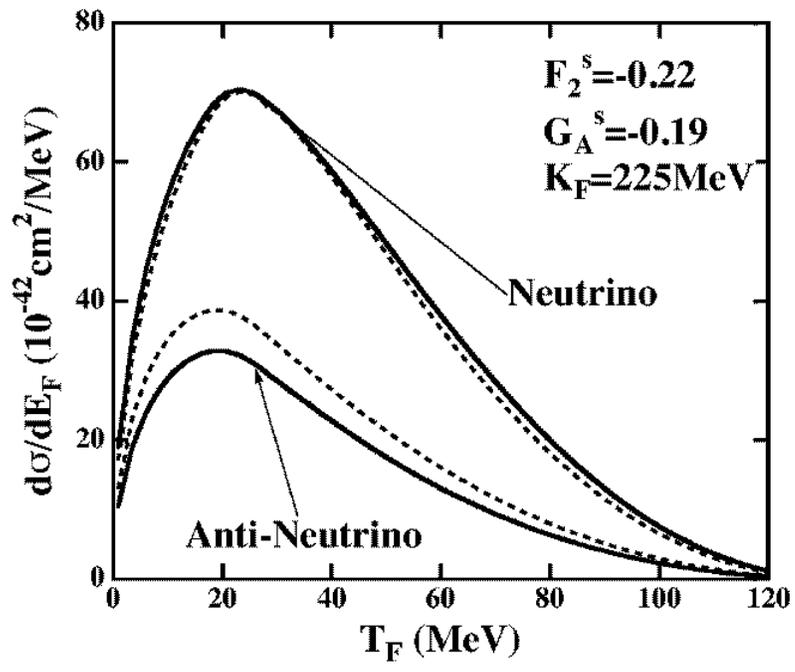
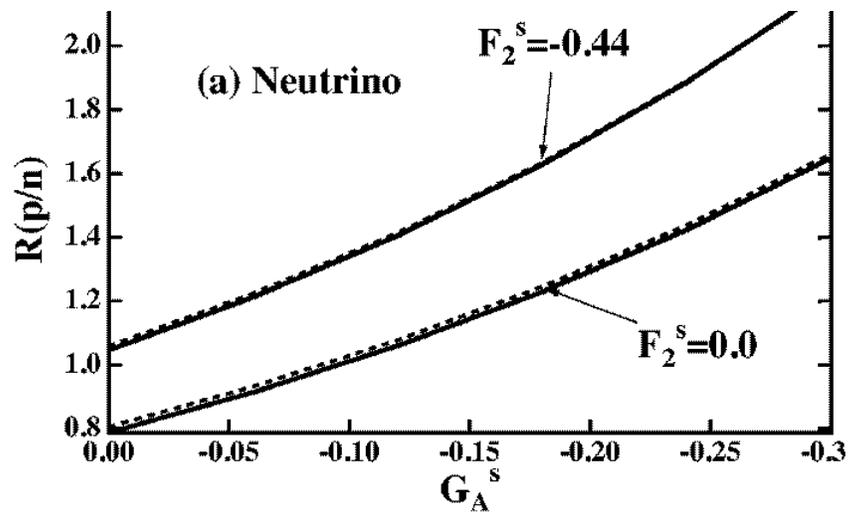
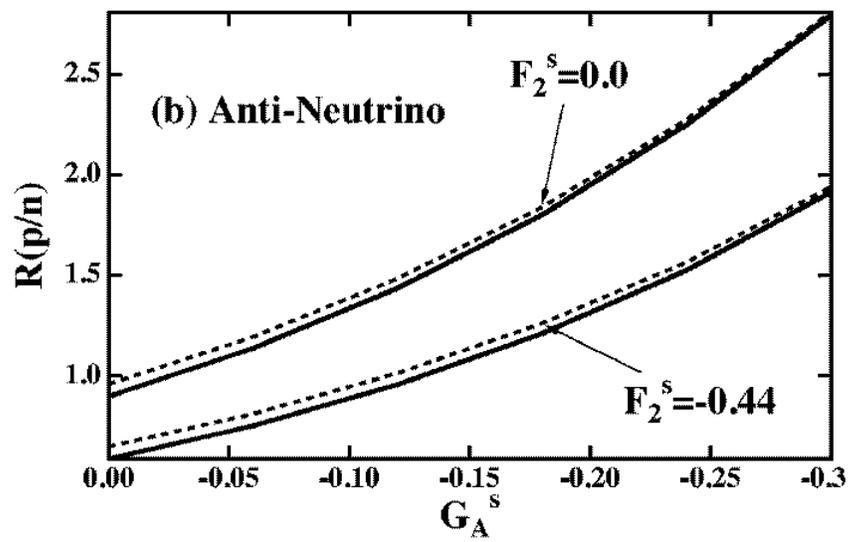
\end{document}